\documentclass[aps,pra,reprint,groupedaddress,twocolumn,10pt]{revtex4-1}
\usepackage{gnuplottex}
\usepackage{mathtools}
\usepackage{lipsum}
\usepackage[usenames,dvipsnames]{xcolor}
\usepackage{float}

\newcommand{\etal}{{\it{et al.}}}
\newcommand{\subsupi}[3]{#1_{\rm #2}^{\rm #3}}

\definecolor{dgreen}{rgb}{0,.5,0}
\definecolor{dblue}{rgb}{0,0,0.75}
\definecolor{dred}{rgb}{0.5,0,.5}

\newcommand{\alam}[1]{\textcolor{dblue}{#1}}

\DeclareMathOperator*{\argmin}{arg\,min}

\begin{document}

\preprint{}

\title{Ghost interaction correction in 
ensemble density-functional theory for excited states with and without
range separation}


\author{
Md. Mehboob Alam$^{1{\ast}}$, Stefan
Knecht$^2$, and Emmanuel Fromager$^{1}$\\
\footnote{$^\ast$Corresponding author. Email: malam@unistra.fr}
{\em{
$^1$Laboratoire de Chimie Quantique,
Institut de Chimie, CNRS/Universit\'{e} de Strasbourg,\\
4 rue Blaise Pascal, 67000 Strasbourg, France\\
\vspace{0.3cm}
$^2$Laboratory of Physical Chemistry, ETH Z\"urich,\\
Vladimir-Prelog Weg 2, CH-8093 Z{\"u}rich, Switzerland
}}
}


\date{\today}

\begin{abstract}


Ensemble density-functional theory (eDFT) suffers from the so-called
``ghost interaction" error when approximate exchange-correlation
functionals are used. In this work, we present a rigorous 
ghost interaction
correction (GIC) scheme in the context of range-separated 
eDFT. The method relies on an exact decomposition of the
ensemble short-range exchange-correlation energy into a
multideterminantal exact exchange term, which involves the long-range
interacting ensemble density matrix instead of the Kohn--Sham (KS) one, and a complementary
density-functional correlation energy. A generalized adiabatic
connection formula is derived for the latter.  
In order to perform practical calculations, the
complementary correlation functional has been simply modeled by its ground-state
local density approximation (LDA) while long-range interacting ground- and
excited-state wavefunctions have been obtained self-consistently by combining a long-range
configuration interaction calculation with a short-range LDA potential.    
We show that GIC reduces the curvature of approximate range-separated ensemble 
energies drastically while providing considerably more accurate excitation 
energies, even for charge-transfer and double excitations.
Interestingly, the method performs well also in the context of standard
KS-eDFT, which is recovered when the range-separation parameter
is set to zero. 


\end{abstract}



\maketitle


\section{Introduction}
The low computational cost and good accuracy of time-dependent
density-functional theory
(TD-DFT)~\cite{Casida_tddft_review_2012,marques2004time} 
has made it one of the most popular method for calculating electronic 
excitation energies.
 { Nevertheless, because of the wrong 
asymptotic behavior of approximate density-functional exchange-correlation potentials used in 
TDDFT, it suffers from limitations like the poor description of 
charge-transfer and Rydberg excitations ~\cite{Casida_tddft_review_2012}. 
Additionally, because of the standard adiabatic approximation ({\it i.e.} the use 
of a frequency-independent kernel), the excitations of multiple 
character~\cite{maitra2004double} are completely absent from the
spectrum.}
{
{The present work}} deals with ensemble DFT (eDFT)
~\cite{JPC79_Theophilou_equi-ensembles,PRA_GOK_RRprinc,PRA_GOK_EKSDFT,GOK3}
which is a time-independent alternative to TD-DFT for excited states.
Its variational nature and hence the ease of implementation have caused
its recent reappearance in the
literature~\cite{PRA13_Pernal_srEDFT,ac_fromager,Burke_ensemble,yang2014exact,JCP14_Filatov_conical_inter_REKS,pastorczak2014ensemble,filatov2015ensemble,Filatov-2015-Wiley,pernal2016ghost,ac_fromager}. Originally formulated by Theophilou for
equiensembles~\cite{JPC79_Theophilou_equi-ensembles}, it was 
generalized by Gross {\etal}~\cite{PRA_GOK_RRprinc,PRA_GOK_EKSDFT,GOK3}
about three decades ago but till now it has not gained the status of
a standard method. One of the main reasons is the absence of reliable
exchange-correlation functionals for ensembles whose development remains
challenging~\cite{Nagy_enseXpot,ParagiXens,ParagiXCens,pernal2015excitation}.
Employing ground-state local or semi-local functionals in practical
eDFT calculations usually gives curved ensemble
energies~\cite{senjean2015linear} and introduces so-called "ghost
interaction" errors~\cite{ensemble_ghost_interaction}. {
The latter are induced by 
unphysical interactions between ground and excited states} that 
appear when the Hartree energy is calculated
with an ensemble density ({\it i.e.} a weighted sum of individual state
densities). In spite of these difficulties, the ability of eDFT to account for multiple
excitations~\cite{senjean2015linear}, in particular, motivated 
recent developments, including its multi-configurational extension~\cite{ac_fromager,PRA13_Pernal_srEDFT}.
Very recently, Pernal and coworkers~\cite{PRA13_Pernal_srEDFT}
introduced range separation in eDFT. In their approach, Boltzmann
ensemble weights are defined by means of an effective temperature
parameter that can be tuned, in addition to the range separation
parameter. In contrast, Senjean {\etal}~\cite{senjean2015linear,extrapol_edft} use a linear
interpolation method (LIM) in order to obtain weight-independent excitation
energies. Since LIM uses ensemble densities in conjunction with
ground-state Hartree-exchange-correlation (Hxc) functionals, it
obviously suffers from ghost interaction errors. In Pernal's scheme,
the error is pragmatically removed 
by defining individual state energies~\cite{pernal2016ghost}. So far,
rigorous ghost interaction corrections have been developed in the
context of single determinantal Kohn--Sham (KS)
eDFT~\cite{ensemble_ghost_interaction,tasnadi_nagy2003,pastorczak2014ensemble}
only. In this work, we present a rigorous strategy for removing ghost
interaction errors in 
range-separated eDFT which, at the end, proves to be equally applicable to standard KS-eDFT.

The paper is organized as follows: After a brief review on exact
range-separated eDFT (Sec.~\ref{subsec:sreDFT}) and an introduction to the usual weight-independent density-functional
approximation (Sec.~\ref{widfa}), the concept of ghost interaction as
well as an exact ghost-interaction-free expression for the range-separated ensemble
energy are presented in Sec.~\ref{subsec:GIC_mdEXX}. Approximate
implementable formulations with and without extrapolation corrections
are then provided in Sec.~\ref{subsec:GIC-extrapol}. Following the
computational details (Sec.~\ref{sec:comp_details}), numerical results are
discussed in Sec.~\ref{sec:results_discussion}. Conclusions are given in
Sec.~\ref{sec:conclu}.
      
\section{Theory}

\subsection{Range-separated ensemble density-functional theory for
excited states}\label{subsec:sreDFT}

In eDFT, an ensemble consisting of $M$ eigenstates
$\{{\Psi}_{k}[v] \} _{0\leq k\leq M-1}$ of the Hamiltonian 
$\hat{H}[v] =
\hat{T} + \hat{W}_{\rm{ee}} +\int d{\bf r}\ v({\bf r})\hat{n}({\bf r})
$
with energies
$E_{0}[v],E_{1}[v],\dots,E_{M-1}[v]$ and the associated weights
{\bf{w}}$\equiv \left(w_{0},w_{1},\dots ,w_{M-1}\right)$ is considered. 
The operators $\hat{T}$, $\hat{W}_{\rm{ee}}$, and $\hat{n}({\bf r})$
correspond to the kinetic
energy, the regular two-electron
repulsion and the density, respectively. {The weights are assigned in such a way that} $w_{0} \geq w_{1} \geq \dots \geq w_{M-1}$ and
$\sum_{k=0}^{M-1}w_{k}=1$. According to the Gross--Oliveira--Kohn (GOK)
variational principle~\cite{PRA_GOK_RRprinc}, the following inequality
holds for any trial ensemble density matrix
$\hat{\gamma}^{\bf{w}} =
\sum_{k=0}^{M-1}w_{k}\vert\overline{\Psi}_{k}
\rangle\langle\overline{\Psi}_{k} \vert$, 
\begin{eqnarray}
\begin{array}{l}\label{eq:GOK_variational}
E^{\rm{\bf{w}}}[v] \leq \text{Tr}\left[ \hat{\gamma}^{\rm{\bf{w}}}\hat{H}[v]\right],
\end{array}
\end{eqnarray}
where Tr denotes the trace. The lower
bound,  
\begin{eqnarray}\label{eq:EE_for_v}
E^{\rm{\bf{w}}}[v] &=&\text{Tr}\left[
\hat{\Gamma}^{\rm{\bf{w}}}[v]\hat{H}[v]\right]
\nonumber\\
&=&\sum_{k=0}^{M-1}w_{k}E_{k}[v], 
\end{eqnarray}
is the exact
ensemble energy that is reached when the trial density matrix equals the exact ensemble density matrix 
$\hat{\Gamma}^{\rm{\bf{w}}}[v] = \sum_{k=0}^{M-1}w_{k}\vert{\Psi}_{k}[v]
\rangle\langle {\Psi}_{k}[v] \vert$. 
An important consequence of this variational principle is that the Hohenberg--Kohn
theorem can be extended to ensembles~\cite{PRA_GOK_EKSDFT}, thus leading
to the exact variational expression 
{
\begin{eqnarray}
\label{eq:EDFT_energy}
E^{\rm{\bf{w}}}[v] = \min\limits_{n}\left\{ F^{\rm{\bf{w}}}[n] + \int d{\bf{r}}\ v({\bf{r}})n({\bf{r}})\right\},
\end{eqnarray}
where 
\begin{eqnarray}\label{eq:LL_ens_func}
F^{\rm{\bf{w}}}[n]&=&
\min\limits_{\hat{\gamma}^{{\bf{w}}}\rightarrow n} 
\text{Tr}\left[
\hat{\gamma}^{\bf{w}}( \hat{T} + \hat{W}_{\rm{ee}} ) \right]
\nonumber\\
&=&
\text{Tr}\left[
\hat{\Gamma}^{\bf{w}}[n]( \hat{T} + \hat{W}_{\rm{ee}} ) \right]
\end{eqnarray} is the
analogue of the Levy--Lieb (LL) functional for ensembles. 
Note that the minimization in Eq.~(\ref{eq:LL_ens_func}) is performed
over all ensemble density matrices with density $n({\bf r})$:
\begin{eqnarray}
\hat{\gamma}^{{\bf{w}}}\rightarrow n \Leftrightarrow
\text{Tr}\left[\hat{\gamma}^{{\bf{w}}}\hat{n}({\bf r})\right]
=n_{\hat{\gamma}^{{\bf{w}}}}({\bf r}) =n({\bf r}).
\end{eqnarray}
Note also that, for any trial density $n({\bf r})$, the GOK inequality in
Eq.~(\ref{eq:GOK_variational}) can be
applied to the minimizing ensemble density matrix
$\hat{\Gamma}^{\bf{w}}[n]$ with density $n({\bf r})$, thus leading to
\begin{eqnarray}
\label{eq:GOK_variational_LF1}
E^{\rm{\bf{w}}}[v] \leq \text{Tr}\left[ \hat{\Gamma}^{\rm{\bf{w}}}[n]\hat{H}[v]\right],
\end{eqnarray}
or, equivalently, according to Eq.~(\ref{eq:LL_ens_func}), 
\begin{eqnarray}
\label{eq:GOK_variational_LF2}
E^{\rm{\bf{w}}}[v] \leq F^{\rm{\bf{w}}}[n]+
\int d{\bf{r}}\,v({\bf{r}})n({\bf{r}})
.  
\end{eqnarray}
Since Eq.~(\ref{eq:GOK_variational_LF2}) holds for any potential $v({\bf
r})$, $F^{\rm{\bf{w}}}[n]$ can be rewritten as a Legendre--Fenchel
transform, exactly like in the ground-state theory~\cite{LFTransform-Lieb}:
\begin{eqnarray}
\label{eq:legendre_fenchel_ensembleLL}
F^{\bf w}[n] = \sup\limits_{v} \left\{ E^{\bf w}[v] 
-\int d{\bf{r}}\,v({\bf{r}})n({\bf{r}})
\right\}.
\end{eqnarray}
From a mathematical point of view, the latter expression is well defined
since the ensemble energy, in contrast to individual excited-state
energies, is concave with respect to the local potential $v({\bf{r}})$.
Indeed, for any potentials $v_{a}({\bf r})$ and $v_{b}({\bf r})$, and
any $\zeta$ in the range $0\leq \zeta\leq 1$, the exact ensemble
energy associated with the average potential $v^\zeta({\bf r})=(1-\zeta)v_{a}({\bf r}) + \zeta v_{b}({\bf r})$
reads (see Eq.~(\ref{eq:EE_for_v})) 
\begin{eqnarray}\label{eq:concavity_ens_energy_atv1}
E^{\bf w}\left[v^\zeta\right] 
&=& 
\text{Tr}\left[ \hat{\Gamma}^{\bf w}[v^\zeta]
\hat{H}[v^\zeta]\right]
\nonumber\\
&=&(1-\zeta)
\text{Tr}\left[ \hat{\Gamma}^{\bf w}[v^\zeta]
\hat{H}[v_a]\right]
\nonumber\\
&&+\zeta
\text{Tr}\left[ \hat{\Gamma}^{\bf w}[v^\zeta]
\hat{H}[v_b]\right].
\end{eqnarray}
Therefore, applying the GOK principle to both $\hat{H}[v_a]$ and
$\hat{H}[v_b]$ Hamiltonians leads to the concavity relation,
\begin{eqnarray}\label{eq:concavity_ens_energy_atv2}
E^{\bf w}\left[v^\zeta\right] 
&\geq&
(1-\zeta)E^{\bf w}\left[ v_{a} \right] + \zeta E^{\bf w}\left[ v_{b}
\right].
\end{eqnarray}
Finally, like in the ground-state theory, differentiability problems
of the ensemble LL functional should in principle occur in directions that change the number of
electrons. It was shown recently by Helgaker and
coworkers~\cite{differentiable2014} that a differentiable but exact
formulation of DFT can be obtained by using a Moreau-Yosida
regularization. It would actually be interesting to explore the
extension of this work to eDFT.\\

Returning to the main focus of this paper, which is the
ghost interaction problem in range-separated eDFT,
we decompose the
two-electron interaction into long- and short-range contributions~\cite{savinstoll,savinbook,savin1988combined}, 
\begin{eqnarray}
\begin{array}{l}\label{eq:range_separation}
\subsupi{\hat{W}}{ee}{} = \subsupi{\hat{W}}{ee}{lr,\mu}  +\subsupi{\hat{W}}{ee}{sr,\mu},\ \ \
\subsupi{\hat{W}}{ee}{lr,\mu} \equiv \sum\limits_{i < j} \frac{{\rm erf}(\mu \vert {\bf r}_{i} - {\bf{r}}_{j} \vert)}{\vert {\bf r}_{i} - {\bf{r}}_{j} \vert},
\end{array}
\end{eqnarray}
where erf is the error function and $\mu$ is a parameter in
$[0,+\infty[$ that controls
the range separation. According to Eq.~(\ref{eq:range_separation}), the
ensemble LL functional can be range-separated as follows, 
\begin{eqnarray}
\begin{array}{l}\label{eq:ens_LL_functional}
F^{\rm{\bf{w}}}[n] = F^{\rm{lr,\mu,\bf{w}}}[n] + E_{\rm{Hxc}}^{\rm{sr,\mu,{\bf{w}}}}[n],
\end{array}
\end{eqnarray}
where, by analogy with Eq.~(\ref{eq:LL_ens_func}), 
\begin{eqnarray}
\begin{array}{l}\label{eq:lr_LL_functional}
\begin{split}
F^{\rm{lr,\mu,{\bf{w}}}}[n] & = \min\limits_{\hat{\gamma}^{\rm{\bf{w}}}\rightarrow n}
\left\{\text{Tr}[\hat{\gamma}^{\bf{w}}(\hat{T} + \hat{W}_{\rm{ee}}^{\rm{lr,\mu}}) ] \right\}\\
& = \text{Tr}\left[\hat{\Gamma}^{\mu,{\bf{w}}}[n](\hat{T} +
\hat{W}_{\rm{ee}}^{\rm{lr,\mu}}) \right],
\end{split}
\end{array}
\end{eqnarray}
and $E_{\rm{Hxc}}^{\rm{sr,\mu,{\bf{w}}}}[n]$ is the complementary
short-range ensemble Hxc functional which
is both ${\bf{w}}$- and $\mu$-dependent. 
Note that $\hat{\Gamma}^{\mu,{\bf{w}}}[n]$ is the density matrix of the
long-range-interacting ensemble with density $n$. The short-range
ensemble Hxc energy is
usually split as
follows~\cite{PRA13_Pernal_srEDFT,ac_fromager,senjean2015linear},
\begin{eqnarray}
\begin{array}{l}\label{eq:ens_Hxc_partition}
E_{\rm{Hxc}}^{\rm{sr,\mu,{\bf{w}}}}[n] = E_{\rm{H}}^{\rm{sr,\mu}}[n] + E_{\rm{xc}}^{\rm{sr,\mu,{\bf{w}}}}[n],
\end{array}
\end{eqnarray}
where the (weight-independent) short-range Hartree term equals 
\begin{eqnarray}
\begin{array}{l}\label{eq:srEDFT_hartree}
\displaystyle
\subsupi{E}{H}{sr,\mu}[n] = \frac{1}{2}\int\int d{\bf r}\ d{\bf{r}^{\prime}}n({\bf r})n({\bf{r}}^{\prime})
\frac{{\rm{erfc}}\left( \mu \vert {\bf r} - {\bf{r}^{\prime}} \vert\right)}{\vert {\bf r} - {\bf{r}^{\prime}} \vert},
\end{array}
\end{eqnarray}
with ${\rm{erfc}}(x) = 1 - {\rm{erf}}(x)$. For a given electronic system
with a nuclear potential $v_{\rm{ne}}({\bf r})$, combining Eq.~(\ref{eq:EDFT_energy}) with Eqs.~(\ref{eq:ens_LL_functional}) and ~(\ref{eq:lr_LL_functional})
leads to the following variational range-separated expression for the exact ensemble
energy $E^{\bf{w}}[v_{\rm{ne}}]$ that we simply denote $E^{\bf{w}}$ in
the following~\cite{senjean2015linear},
\begin{align}\label{eq:srEDFT_ens_energy}
E^{\bf{w}}  & = \min_{\hat{\gamma}^{\bf{w}}}\left\{ \text{Tr}\left[ 
\hat{\gamma}^{\rm{{\bf{w}}}} ( \hat{T} + \hat{W}_{\rm{ee}}^{\rm{lr,\mu}} + \hat{V}_{\rm{ne}}) \right] +
E_{\rm{Hxc}}^{\rm{sr,\mu,{\bf{w}}}}[n_{\hat{\gamma}^{\rm{{\bf{w}}}}}] \right\} \nonumber \\ 
 & = \text{Tr}[ 
\hat{\Gamma}^{\rm{\mu,{\bf{w}}}} ( \hat{T} + \hat{W}_{\rm{ee}}^{\rm{lr,\mu}} + \hat{V}_{\rm{ne}}) ] +
E_{\rm{Hxc}}^{\rm{sr,\mu,{\bf{w}}}}[n_{\hat{\Gamma}^{\rm{\mu,{\bf{w}}}}}],
\end{align}
where $\hat{V}_{\rm{ne}}=\int d{\bf r}\,v_{\rm{ne}}({\bf r})\hat{n}({\bf
r})$. The minimizing density matrix
$\hat{\Gamma}^{\rm{\mu,{\bf{w}}}}=
\sum_{k=0}^{M-1}w_{k}\vert\Psi_{k}^{\rm{\mu,{\bf{w}}}}\rangle\langle\Psi_{k}^{\rm{\mu,{\bf{w}}}}\vert$
reproduces the exact physical ensemble density,
$n_{\hat{\Gamma}^{\rm{\mu,{\bf{w}}}}}({\bf{r}}) =
\text{Tr}[\hat{\Gamma}^{\rm{\mu,{\bf w}}}\hat{n}({\bf r})] =
n_{\hat{\Gamma}^{\rm{\bf{w}}}[v_{\rm{ne}}]}({\bf{r}})$, and the corresponding
wavefunctions $\{\Psi_{k}^{\rm{\mu,{\bf{w}}}}\}_{0\leq k\leq M-1}$
fulfill the following self-consistent
equations~\cite{senjean2015linear}, 
\begin{multline}\label{eq:srEDFT_scf_eq}
\left( \hat{T} + \hat{W}_{\rm{ee}}^{\rm{lr,\mu}} + \hat{V}_{\rm{ne}} + 
\int\ d{\bf{r}}\frac{\delta E_{\rm{Hxc}}^{\rm{sr,\mu,{\bf{w}}}}[n_{\hat{\Gamma}^{\rm{\mu,{\bf{w}}}}}]}
{\delta n({\bf{r})}} \hat{n}({\bf{r}}) \right) \vert\Psi_{k}^{\rm{\mu,{\bf{w}}}} \rangle \\
= \mathcal{E}_{k}^{\rm{\mu,{\bf{w}}}} \vert\Psi_{k}^{\rm{\mu,{\bf{w}}}} \rangle, 0 \leq k \leq M-1.
\end{multline}
Note that the standard Schr\"{o}dinger and KS-eDFT equations are
recovered from Eq.~(\ref{eq:srEDFT_scf_eq}) for $\mu \rightarrow
+\infty$ and $\mu=0$, respectively. For the sake of simplicity, we will focus in the following on
two-state ensembles. In this particular case, one single weight $w$ with
$0 \leq w \leq 1/2$ is needed and ${\bf{w}}\equiv(1-w,w)$, so that the
exact ensemble energy reads
\begin{eqnarray}
E^{w} = (1- w)E_{0} + wE_{1}, 
\end{eqnarray}
where $E_{k}=E_{k}[v_{\rm{ne}}]$, $k=0,1$.
Let us stress that all the methods discussed in the following can be
extended straightfowardly to higher excitations simply by considering larger
ensembles and expressing the targeted excitation energy in terms of
equiensemble energies and lower excitation
energies~\cite{PRA_GOK_EKSDFT,senjean2015linear}. This will be
discussed in more details in Sec.~\ref{widfa}.
\\

In recent works, Senjean {\etal}~\cite{senjean2015linear,extrapol_edft}  
pointed out that, in the exact theory, the excitation energy
can be calculated in (at least) two different ways. The first one consists in
differentiating the ensemble energy in Eq.~(\ref{eq:srEDFT_ens_energy})
with respect to the ensemble weight, thus leading to~\cite{senjean2015linear}
\begin{eqnarray}
\begin{array}{l}\label{eq:deri_exact_ExEn}
\begin{split}
 \omega & = \frac{dE^{\it w}}{d{\it w}} = \mathcal{E}_{1}^{\mu,w} - \mathcal{E}_{0}^{\mu,w} 
 + \left.\dfrac{\partial \subsupi{E}{Hxc}{sr,\mu , {\it w}}[n]}
 {\partial w}\right\vert_{n=n_{\hat{\Gamma}^{\mu,\it w}}} \\
& = \Delta\subsupi{\mathcal{E}}{}{\mu , {\it w}} + \subsupi{\Delta}{xc}{\mu , {\it w}},
\end{split}
\end{array}
\end{eqnarray}
where $\Delta\subsupi{\mathcal{E}}{}{\mu , {\it
w}}=\mathcal{E}_{1}^{\mu,w} - \mathcal{E}_{0}^{\mu,w}$ is the auxiliary
long-range-interacting excitation energy and $\subsupi{\Delta}{xc}{\mu ,
{\it w}}$ is the short-range analogue of the xc derivative discontinuity
for a canonical ensemble~\cite{PRA_Levy_XE-N-N-1,ac_fromager}. 
In the $\mu=0$ limit, this derivative with respect to the ensemble weight $w$ corresponds, when
$w=0$, to the jump in the KS HOMO energy that occurs when
comparing $w=0$ and $w\rightarrow 0$ situations, hence the name
"derivative discontinuity". This was shown by
Levy~\cite{PRA_Levy_XE-N-N-1} and observed numerically by Yang~{\it et
al.}~\cite{yang2014exact} in the He atom. The proof is very
similar to the one for the discontinuity due to the change of particle
number but the two discontinuities are different. Indeed, we consider
here a canonical ensemble where both ground and excited states
have the same number of electrons.\\

In an alternative approach,
referred to as LIM~\cite{senjean2015linear}, the exact excitation
energy is simply obtained by linear interpolation, 
\begin{eqnarray}
\begin{array}{l}\label{eq:exactLIM}
\omega = 2 \left( E^{w=1/2} - E^{w=0}\right),
\end{array}
\end{eqnarray}
where $E^{w=0}=E_0$ is the exact ground-state energy.

\subsection{Weight-independent density-functional approximation
}\label{widfa}

Let us stress that Eqs.~(\ref{eq:deri_exact_ExEn}) 
and (\ref{eq:exactLIM}) are equivalent if exact functionals
and wave functions are used, which is of course not the case in practical
calculations~\cite{senjean2015linear}. In the standard
weight-independent density functional approximation
(WIDFA)~\cite{PRA13_Pernal_srEDFT,senjean2015linear,extrapol_edft}, the
ensemble energy in Eq.~(\ref{eq:srEDFT_ens_energy}) and the auxiliary
wave functions in Eq.~(\ref{eq:srEDFT_scf_eq}) are calculated by
substituting the short-range ensemble functional with the (weight-independent) ground-state
one $\subsupi{E}{Hxc}{sr,\mu}[n]={E}_{\rm Hxc}^{{\rm sr},\mu,w=0}[n]$, thus leading to the approximate WIDFA
variational ensemble energy,
\begin{eqnarray}\label{eq:widfa_ens_en}
\tilde{E}^{\rm{\mu,{\it w}}} &=&
\min_{\hat{\gamma}^{{w}}}\left\{ \text{Tr}\left[ 
\hat{\gamma}^{{{{w}}}} ( \hat{T} + \hat{W}_{\rm{ee}}^{\rm{lr,\mu}} + \hat{V}_{\rm{ne}}) \right] +
E_{\rm{Hxc}}^{\rm{sr,\mu}}[n_{\hat{\gamma}^{{{{w}}}}}] \right\} \nonumber \\ 
\nonumber\\
&=& \text{Tr}[ 
\hat{\gamma}^{\rm{\mu,{\it{w}}}} ( \hat{T} + \hat{W}_{\rm{ee}}^{\rm{lr,\mu}} + \hat{V}_{\rm{ne}}) ] +
E_{\rm{Hxc}}^{\rm{sr,\mu}}[n_{\hat{\gamma}^{\rm{\mu,{\it{w}}}}}],
\end{eqnarray}
the corresponding WIDFA ensemble density matrix, 
\begin{eqnarray}\label{eq:widfa_conditions}
\hat{\gamma}^{\mu,w}&=&
 (1-w)
\vert\tilde{\Psi}^{\mu,w}_0\rangle\langle
\tilde{\Psi}^{\mu,w}_0\vert
+w
\vert\tilde{\Psi}^{\mu,w}_1\rangle\langle
\tilde{\Psi}^{\mu,w}_1\vert
,
\end{eqnarray}
and, according to Eqs.~(\ref{eq:deri_exact_ExEn}) and (\ref{eq:exactLIM}), 
to the weight- and $\mu$-dependent excitation energy expression,
\begin{eqnarray}\label{eq:widfaXE}
\begin{array}{l}
\omega \rightarrow \Delta\tilde{\mathcal{E}}^{\mu,{\it w}}=
\tilde{\mathcal{E}}_{1}^{\mu,w} - \tilde{\mathcal{E}}_{0}^{\mu,w},
\end{array}
\end{eqnarray}
or, alternatively, to 
\begin{eqnarray}
\begin{array}{l}\label{eq:widfa_LIM}
\begin{split}
\omega \rightarrow \tilde{\omega}^\mu_{\rm{LIM}} = 2\left( \tilde{E}^{\mu,w=1/2} -
\tilde{E}^{\mu,w=0}\right).
\end{split}
\end{array}
\end{eqnarray}
The latter expression is, by construction, weight-independent. It only
depends on the $\mu$ parameter. Note that the ground-state energy
$\tilde{E}^{\mu,w=0}$ will be $\mu$-dependent in practice since approximate
ground-state functionals are used. Let us emphasize that 
Eq.~(\ref{eq:widfa_LIM}) can be extended to higher excitations and degenerate states through linear interpolations
between equiensembles~\cite{senjean2015linear}, thus leading to the
following expression for the $I$th excitation energy,
\begin{eqnarray}\label{eq:generalizedWIDFA_LIM_exen}
\tilde{\omega}_{{\rm{LIM,}}I}^{\rm{\mu}} &=& \frac{M_{I}}{g_{I}}
\left( \tilde{E}_{I}^{{\rm{\mu}},1/M_{I}} - \tilde{E}_{I-1}^{{\rm{\mu}},1/M_{I-1}}\right)
\nonumber\\
&&+ \frac{1}{M_{I-1}}\sum\limits_{k=1}^{I-1}g_{k}\tilde{\omega}^\mu_{{\rm{LIM}},k},
\end{eqnarray}
where $g_{k}$ is the degeneracy of the $k$th energy, $M_{I} =
\sum_{k=0}^{I}g_{k}$ is the total number of states in the targeted
equiensemble (the one that enables to reach the $I$th energy) and
$\tilde{E}_{I}^{{\rm{\mu}},1/M_{I}}$ is the corresponding WIDFA equiensemble energy
(with weight $1/M_{I}$). Note that each equiensemble is made of multiplets.
In other words, all degenerate states should be included.\\

In the formulation of range-separated eDFT by
Pastorczak {\it et al.}~\cite{PRA13_Pernal_srEDFT}, the WIDFA is also
used but excitation energies are computed differently. A single ensemble
containing all states of interest is calculated (from
Eq.~(\ref{eq:srEDFT_scf_eq}) with
the substitution $E_{\rm{Hxc}}^{\rm{sr,\mu,{\bf{w}}}}[n]\rightarrow
E_{\rm{Hxc}}^{\rm{sr,\mu}}[n]$) and individual state
energies are pragmatically introduced as follows,   \\ 
\begin{eqnarray}\label{eq:pernal_energies}
\tilde{E}^{\mu,{\bf{w}}}_k&=&\langle\tilde{\Psi}_k^{\mu,{\bf{w}}}\vert\hat{T}
+ \hat{W}_{\rm{ee}}^{\rm{lr,\mu}} + \hat{V}_{\rm{ne}}\vert\tilde{\Psi}_k^{\mu,{\bf{w}}}\rangle
\nonumber\\
&&+E_{\rm{Hxc}}^{\rm{sr,\mu}}[n_{\tilde{\Psi}_k^{\mu,{\bf{w}}}}].
\end{eqnarray}
As discussed in Ref.~\cite{senjean2015linear}, the latter expression is
questionable, especially because it uses individual state densities
(rather than the ensemble density) in conjunction with the ground-state
short-range functional. Let us stress that, in contrast to LIM, even if exact  
functionals and wavefunctions were used, the energies in
Eq.~(\ref{eq:pernal_energies}) would {\it not}, in principle, be exact.
This statement holds for any
finite $\mu$ value. A simple argument
is that, for the ground-state energy, the long-range interacting wavefunction
$\tilde{\Psi}_0^{\mu,{\bf{w}}}$ will not have its density equal to the
exact ground-state density of the physical system. The former density
will contribute to a long-range interacting
ensemble density that is equal to the exact ensemble density of the
physical system. Another
practical issue that arises when approximations are made is that the
state energies in Eq.~(\ref{eq:pernal_energies}) and, consequently, the
excitation energies depend on both the range-separation parameter $\mu$
and the ensemble weights ${\bf{w}}$. As Boltzmann weights are used in
the scheme of Pastorczak {\it et al.}~\cite{PRA13_Pernal_srEDFT}, they are all controlled by an
effective inverse temperature $\beta$ which is a tunable parameter in
the theory. In this respect, LIM has the advantage of providing
excitation energies that are, by construction, weight-independent.
Defining approximate excitation energies by linear interpolation is of
course a choice. Others would be possible. 

\subsection{Ghost interaction and alternative range-separated ensemble energy
expression 
}\label{subsec:GIC_mdEXX}

Let us return to the two-state ensemble problem. Although the
combination of LIM and WIDFA  
gave promising results~\cite{senjean2015linear,extrapol_edft}, the use of local or
semi-local ground-state short-range xc functionals inevitably introduces
a so-called "ghost interaction" error~\cite{ensemble_ghost_interaction} in the  
equiensemble energy $\tilde{E}^{\mu,w=1/2}$ and, consequently, in the
LIM excitation energy (see Eqs.~(\ref{eq:widfa_ens_en}) and
(\ref{eq:widfa_LIM})). This error arises when inserting the WIDFA
ensemble density 
\begin{eqnarray}
n_{\hat{\gamma}^{\mu,w}}({\bf r})=(1-w)n_{\tilde{\Psi}^{\mu,w}_0}({\bf
r})+w\,n_{\tilde{\Psi}^{\mu,w}_1}({\bf r})
\end{eqnarray}
into the short-range Hartree term (see
Eqs.~(\ref{eq:ens_Hxc_partition})
and (\ref{eq:srEDFT_hartree})):   
\begin{eqnarray}\label{eq:gi_expression}
 &&E_{\rm{H}}^{\rm{sr,\mu}}[n_{\hat{\gamma}^{\mu,w}}]=
(1-w)^2E_{\rm{H}}^{\rm{sr,\mu}}[n_{\tilde{\Psi}^{\mu,w}_0}]
+w^2E_{\rm{H}}^{\rm{sr,\mu}}[n_{\tilde{\Psi}^{\mu,w}_1}]
\nonumber\\
&&+ w(1-w)\int\int d{\bf r}d{\bf{r^{\prime}}}
n_{\tilde{\Psi}^{\mu,w}_0}({\bf r})n_{\tilde{\Psi}^{\mu,w}_1}({\bf{r^{\prime}}})
\nonumber\\
&&\hspace{2.5cm}\times\frac{\rm{erfc}(\mu\vert {\bf r} 
- {\bf{r^{\prime}}} \vert)}{\vert {\bf r} - {\bf{r^{\prime}}}\vert}.
\end{eqnarray}
As readily seen in Eq.~(\ref{eq:gi_expression}), the last term on the
right-hand side describes an unphysical "ghost interaction" between the
ground and first excited states through the product of their 
densities. This error does not show up in the approach of Pastorczak {\it et
al.}~\cite{PRA13_Pernal_srEDFT,pernal2016ghost} since, as shown in
Eq.~(\ref{eq:pernal_energies}), individual state densities are inserted
into the short-range density functional. As
discussed previously, even though it is convenient, the definition of
individual state energies in the context of eDFT is a pragmatic choice.
In this work, we intend to remove ghost interaction errors in the LIM
excitation energies by applying a correction scheme to the WIDFA
ensemble energy rather than by constructing individual state energies.
For that purpose, we consider the following decomposition of the exact short-range ensemble xc
energy~\cite{extrapol_edft},
\begin{eqnarray}\label{eq:md_xc_decomp}
\subsupi{E}{xc}{sr,\mu,{\it w}}[n]  =
\subsupi{E}{x,md}{sr,\mu,{\it w}}[n] + \subsupi{E}{c,md}{sr,\mu,{\it w}}[n],
\end{eqnarray}
where 
\begin{eqnarray}\label{eq:mdex_expression}
E_{\rm{x,md}}^{\rm{sr,\mu,{\it w}}}[n] =
{\rm Tr}[\hat{\Gamma}^{\mu,{\it w}}[n]\hat{W}^{\rm{sr,\mu}}_{\rm ee}] -
\subsupi{E}{H}{sr,\mu}[n]
\end{eqnarray}
is the analogue of the
multideterminantal (md) short-range exchange functional of Toulouse
{\etal}~\cite{Toulouse2005TCA} for ensembles and
$E_{\rm{c,md}}^{\rm{sr,\mu,{\it w}}}[n]$ is the complementary
short-range ensemble correlation functional. Note that
$\hat{\Gamma}^{\mu,{\it w}}[n]$ is defined in
Eq.~(\ref{eq:lr_LL_functional}) and
corresponds to the long-range interacting ensemble density matrix with
density $n({\bf r})$. Since, according to Eq.~(\ref{eq:srEDFT_scf_eq})
[here we consider the particular case of $M=2$ states] and the
Hohenberg--Kohn theorem for ensembles~\cite{PRA_GOK_EKSDFT},
\begin{eqnarray} 
\hat{\Gamma}^{\mu,{\it w}}[n_{\hat{\Gamma}^{\mu,{\it
w}}}]=\hat{\Gamma}^{\mu,{\it w}},
\end{eqnarray} 
combining Eqs.~(\ref{eq:md_xc_decomp}) and (\ref{eq:mdex_expression}) with
Eqs.~(\ref{eq:range_separation}), (\ref{eq:ens_Hxc_partition}) and
(\ref{eq:srEDFT_ens_energy}) leads to an exact alternative expression
for the range-separated ensemble energy,
\begin{eqnarray}\label{eq:exact_GIC_ener}
E^{\it w} = \text{Tr}[\hat{\Gamma}^{\rm{\mu,{\it w}}}\hat{H}] +
E_{\rm{c,md}}^{\rm{sr,\mu,{\it w}}}[n_{\hat{\Gamma}^{\rm{\mu,{\it
w}}}}],
\end{eqnarray}
where $\hat{H}=\hat{H}[v_{\rm ne}]$ is the true physical Hamiltonian.
Note that, even though the true Hamiltonian (without range separation)
is used, the energy is obtained from a long-range interacting ensemble
density matrix. Therefore, short-range correlation effects are missing
in the first term on the right-hand side of
Eq.~(\ref{eq:exact_GIC_ener}). These effects are described by the
complementary ensemble md short-range correlation functional. 
As readily seen, this alternative energy expression is free from ghost
interaction errors since only short-range correlation effects are now
described with a density functional. Of course, the use of an approximate
correlation functional in this context may introduce residual "ghost
correlation" errors but the numerical results discussed in
Sec.~\ref{sec:results_discussion}
seem to indicate that the latter are not too significant, at least in
the simple two- and four-electron systems considered in this work. Note
that, when $\mu=0$, the ensemble energy expression in
Eq.~(\ref{eq:exact_GIC_ener}) becomes similar to the linear exact
exchange expression of Gould and Dobson for grand canonical ensembles
(see Eq.~(5) in Ref.~\cite{gould2013}).
In order to implement Eq.~(\ref{eq:exact_GIC_ener}) for any $\mu$
values, we need approximate complementary short-range ensemble correlation
functionals. So far, only a ground-state local density approximation
(LDA) has been developed~\cite{Paziani2006PRB}. A simple approximation,
that will be used in Sec.~\ref{sec:results_discussion}, consists in
using the ground-state functional,
\begin{eqnarray}
E_{\rm{c,md}}^{\rm{sr,\mu}}[n]=E_{\rm{c,md}}^{\rm{sr,\mu,{\it w}=0}}[n],
\end{eqnarray} in complete analogy with the WIDFA. In order to get further
insight into what would actually be neglected with such an approximation
and thus pave the way to the construction of adapted weight-dependent
short-range correlation functionals, let us decompose the exact
functional as follows,
\begin{eqnarray}\label{eq:srmdcorr_decomp}
E_{\rm{c,md}}^{\rm{sr,\mu,{\it w}}}[n] = 
E_{\rm{c,md}}^{\rm{sr,\mu}}[n] +
\Delta E_{\rm{c,md}}^{\rm{sr,\mu,{\it w}}}[n],
\end{eqnarray}
where the weight-dependence has been moved to the contribution $\Delta
E_{\rm{c,md}}^{\rm{sr,\mu,{\it w}}}[n]$ for which an adiabatic
connection (AC) formula can be derived. For that purpose, we consider the following 
AC path based on the generalized AC formalism for
ensembles (GACE)~\cite{ac_fromager} and the range-separated AC of
Rebolini~\etal~\cite{MP15_Elisa_PT_XE_AC}:
\begin{align}\label{eq:scf_eq_doubleGACE}
\left(\hat{T} + \hat{W}^{\rm{lr,\mu}}_{\rm{ee}} + \lambda\hat{W}^{\rm{sr,\mu}}_{\rm{ee}} 
+ \hat{V}^{\rm{\mu,\lambda,\xi}} \right) \vert \Psi^{\rm{\mu,\lambda,\xi}}_{k} \rangle \nonumber \\
= \mathcal{E}^{\rm{\mu,\lambda,\xi}}_{k} \vert
\Psi^{\rm{\mu,\lambda,\xi}}_{k} \rangle,\hspace{0.2cm} k = 0,1,
\end{align}
where the local potential 
$\hat{V}^{\rm{\mu,\lambda,\xi}} = \int d{\bf r}\ v^{\rm{\mu,\lambda,\xi}}({\bf r})\ \hat{n}({\bf r})$ 
ensures that the density constraint,
\begin{eqnarray}\label{eq:density_constraint}
\text{Tr}\left[\hat{\Gamma}^{\rm{\mu,\lambda,\xi}}[n]\hat{n}({\bf
r})\right] = n({\bf r}),
\end{eqnarray}
with 
\begin{eqnarray}
\hat{\Gamma}^{\rm{\mu,\lambda,\xi}}[n] &=& 
(1-\xi)\vert \subsupi{\Psi}{0}{\mu,\lambda,\xi} \rangle \langle \subsupi{\Psi}{0}{\mu,\lambda,\xi} \vert 
\nonumber\\
&&+ \xi \vert \subsupi{\Psi}{1}{\mu,\lambda,\xi} \rangle \langle
\subsupi{\Psi}{1}{\mu,\lambda,\xi} \vert,
\end{eqnarray}
is fulfilled not only for all interaction strengths in the range $0
\leq \lambda \leq 1$ but also for all ensemble weights in the range
$0 \leq \xi \leq w$. The constraint is strong and it could potentially lead to
representability problems. Let us mention that in a recent work on the Hubbard dimer (which will be
presented in a separate paper), we have shown that such an AC 
can be constructed. In particular, it appears that
if a density is ensemble representable for a
given weight $w$, then
it is ensemble representable for any weight $\xi$ with 
$0\leq\xi\leq w$. This is a promising result whose extension to the
exact Hamiltonian should be investigated. Work is currently in progress
in this direction.\\ 

Note that the multideterminantal decomposition of the ensemble
short-range xc energy in Eq.~(\ref{eq:md_xc_decomp}) relies on a
fictitious long-range interacting system instead of the usual
non-interacting KS one. Therefore, in order to recover the former system
at $\lambda = 0$, and thus obtain an AC formula for the complementary ensemble short-range
correlation energy, the short-range interaction only is scaled by
$\lambda$ in Eq.~(\ref{eq:scf_eq_doubleGACE}). Therefore, the ensemble density
matrix $\hat{\Gamma}^{\mu,\lambda,\xi}[n]$ reduces to  
$\hat{\Gamma}^{\mu,\xi}[n]$ when $\lambda=0$.
Note that, for $\lambda=1$,
the physical (fully-interacting) system is recovered
($\hat{\Gamma}^{\mu,\lambda=1,\xi}[n]=\hat{\Gamma}^{\xi}[n]$), like in a
conventional AC.
According to Eqs.~(\ref{eq:LL_ens_func}), (\ref{eq:ens_LL_functional}),
(\ref{eq:lr_LL_functional}) and~(\ref{eq:ens_Hxc_partition}), 
the short-range ensemble xc energy can be expressed as 
\begin{eqnarray}
\label{eq:ensX_doubleGACE}
\subsupi{E}{xc}{sr,\mu,{\it w}}[n] = 
\int_{0}^{1} d\lambda \frac{d\subsupi{F}{}{\mu,\lambda,{\it w}}[n]}{d\lambda} 
- \subsupi{E}{H}{sr,\mu}[n],
\end{eqnarray}
where 
\begin{eqnarray}
\subsupi{F}{}{\mu,\lambda,{\it w}}[n] =
\text{Tr}\left[\hat{\Gamma}^{\mu,\lambda,{\it w}}[n](\hat{T} +
\hat{W}^{\rm{lr,\mu}}_{\rm{ee}} +
\lambda\hat{W}^{\rm{sr,\mu}}_{\rm{ee}})\right].
\end{eqnarray}

Using the
Hellmann--Feynman theorem in Eq.~(\ref{eq:ensX_doubleGACE}) with the
density constraint in Eq.~(\ref{eq:density_constraint}) as well as
Eqs.~(\ref{eq:md_xc_decomp}) and (\ref{eq:mdex_expression}) leads to
\begin{eqnarray}
\label{eq:ensX_doubleGACE2}
\subsupi{E}{c,md}{sr,\mu,{\it w}}[n] &=& 
\int_{0}^{1} d\lambda\,
\text{Tr}\left[\hat{\Gamma}^{\mu,\lambda,w}[n]\hat{W}^{\rm{sr,\mu}}_{\rm{ee}}\right] 
\nonumber\\
&&-
\text{Tr}\left[\hat{\Gamma}^{\mu,w}[n]\hat{W}^{\rm{sr,\mu}}_{\rm{ee}}\right]. 
\end{eqnarray}
Finally, from the expression
\begin{eqnarray}
\subsupi{E}{c,md}{sr,\mu,{\it w}}[n]=E_{\rm{c,md}}^{\rm{sr,\mu,{\it
w}=0}}[n]+\int_{0}^{w}d\xi\,\dfrac{\partial
\subsupi{E}{c,md}{sr,\mu,\xi}[n]}{\partial \xi},
\end{eqnarray}
we obtain the decomposition in Eq.~(\ref{eq:srmdcorr_decomp}) with the
following explicit AC formula for the weight-dependent part, 
\begin{eqnarray}
\label{eq:md_c_decomp}
&& \Delta E_{\rm{c,md}}^{\rm{sr,\mu,{\it w}}}[n] = 
\nonumber\\ 
&&\int_{0}^{1}d\lambda\int_{0}^{\it w}d\xi\,
\text{Tr}
 \left[ \left( \frac{\partial\hat{\Gamma}^{\rm{\mu,\lambda,\xi}}[n]}{\partial\xi} 
 - \frac{\partial\hat{\Gamma}^{\rm{\mu,\xi}}[n]}
 {\partial\xi}\right)\hat{W}_{\rm{ee}}^{\rm{sr,\mu}}\right].
\end{eqnarray}

Returning to the energy expression in Eq.~(\ref{eq:exact_GIC_ener}), 
we should stress that, unlike 
the expression in Eq.~(\ref{eq:srEDFT_ens_energy}), it is {\it not} variational with respect to
the ensemble density matrix. Ignoring this leads to double counting
problems~\cite{manusroep2013}, since the minimizing density matrix would
be obtained from a fully-interacting Hamiltonian rather than a
long-range interacting one (as it should). Nevertheless, the ensemble energy
in Eq.~(\ref{eq:exact_GIC_ener}) is variational 
with respect to local potentials. In other words,  
it can be obtain by means of optimized effective potentials
(OEP)~\cite{Toulouse2005TCA} as follows,
\begin{multline}\label{eq:OEP_GIC_energy}
E^{w} = \min\limits_{v} \left\{\text{Tr}[\hat{\Gamma}^{\rm{\mu,{\it w}}}[v] \hat{H} ] 
+ E_{\rm{c,md}}^{\rm{sr,\mu,{\it w}}}[n_{\hat{\Gamma}^{\rm{\mu,{\it w}}}[v]}]\right\},
\end{multline}
where 
\begin{align}\label{eq:OEP_scf_eq}
\hat{\Gamma}^{\rm{\mu,{\it w}}}[v] = \argmin\limits_{\hat{\gamma}^{\it w}}\left\{ \text{Tr}\left[\hat{\gamma}^{\it w} (\hat{T} + \hat{W}_{\rm{ee}}^{\rm{lr,\mu}})\right]\right. \nonumber\\
+ \left.\int d{\bf r}\;v({\bf r})n_{\hat{\gamma}^{\it w}}({\bf r})\right\}.
\end{align}
So far, such a scheme has been implemented efficiently only for
approximate single-determinantal ground-state wave functions but it can,
in principle, be extended to multi-configurational
wave functions~\cite{manusroep2013}. For practical purposes, we will
propose in the following a much simpler
approach where a density-functional potential (the one computed at the
WIDFA level) is used rather than an OEP. In this respect, 
the scheme of Pastorczak {\it et
al.}~\cite{PRA13_Pernal_srEDFT,pernal2016ghost} and the ghost
interaction correction proposed in the following section will be
similar. Both will rely on long-range interacting ensemble density
matrices that are computed similarly from a short-range Hxc
density-functional potential that actually contains ghost interaction
errors (because of the short-range Hartree potential). OEPs would have
the advantage of removing such errors. This is left for future
work.   
\\

Finally, returning to the exact theory and the calculation of the excitation energy, combining
Eq.~(\ref{eq:deri_exact_ExEn}) with Eq.~(\ref{eq:exact_GIC_ener}) leads
to 
\begin{eqnarray}
\begin{array}{l}\label{eq:ExEn_deri_GIC}
\begin{split}
\omega & = \frac{dE^{w}}{dw} = \langle \subsupi{\Psi}{1}{\mu,{\it w}} \vert \hat{H} \vert \subsupi{\Psi}{1}{\mu,{\it w}} \rangle - \langle \subsupi{\Psi}{0}{\mu,{\it w}} \vert \hat{H} \vert \subsupi{\Psi}{0}{\mu,{\it w}} \rangle \\
& + \frac{d \subsupi{E}{c,md}{sr,\mu,{\it w}}[n_{\hat{\Gamma}^{\mu,{\it w}}}]}{d {\it w}}
+ 2w \left. \left.\left\langle \frac{\partial \subsupi{\Psi}{1}{\mu,{\it w}}}{\partial w} \right\vert \hat{H} \right\vert \subsupi{\Psi}{1}{\mu,{\it w}} \right\rangle \\
& + 2(1-w)\left. \left. \left\langle \frac{\partial \subsupi{\Psi}{0}{\mu,{\it w}}}{\partial w} \right\vert \hat{H} \right\vert \subsupi{\Psi}{0}{\mu,{\it w}}\right\rangle.
\end{split}
\end{array}
\end{eqnarray}
Note that the Hellmann--Feynman theorem does not hold because of the
non-variational character (with respect to the ensemble density matrix) of the ensemble energy expression in Eq.~(\ref{eq:exact_GIC_ener}).
 As a
result, the response of both ground- and excited-state wave functions to
variations in the ensemble weight is
in principle needed.

\subsection{Ghost interaction correction and extrapolation
schemes}\label{subsec:GIC-extrapol}

 In order to perform practical excitation energy calculations from
Eq.~(\ref{eq:ExEn_deri_GIC}), we
will consider the following approximations: (i) The
long-range interacting density matrix is calculated at the 
WIDFA level (see Eq.~(\ref{eq:widfa_ens_en})), for example within the
short-range LDA~\cite{savinbook,toulda}.  
(ii) We then use, as an additional approximation and by analogy with WIDFA, the (weight-independent) 
{\it ground-state} functional $\subsupi{E}{c,md}{sr,\mu}[n]$. So far, only an
LDA-type functional has been developed by Paziani {\etal}~\cite{Paziani2006PRB}. If, in addition,
(iii) we neglect the response of both
the ensemble density and the individual
wave functions to variations in $w$, then the approximation (ii) has no
impact on the excitation energy which reduces to a first-order
corrected (FOC) expression~\cite{MP15_Elisa_PT_XE_AC}:
\begin{eqnarray}\label{eq:full-range-corrected_XE}
\begin{array}{l}
\omega \rightarrow  \tilde{\omega}_{\rm FOC}^{\mu,w}=
\langle \subsupi{\tilde{\Psi}}{1}{\mu,{\it w}} \vert \hat{H} \vert \subsupi{\tilde{\Psi}}{1}{\mu,{\it w}} \rangle 
- \langle \subsupi{\tilde{\Psi}}{0}{\mu,{\it w}} \vert \hat{H} \vert \subsupi{\tilde{\Psi}}{0}{\mu,{\it w}} \rangle.
\end{array}
\end{eqnarray}
Note that the latter expression becomes exact only in the
$\mu\rightarrow+\infty$ limit and it converges as $\mu^{-4}$~\cite{savin2014towards,MP15_Elisa_PT_XE_AC}.\\

In order to preserve the
ghost-interaction-free character of the FOC excitation energy while
taking into account the missing short-range
correlation effects, it is in fact simpler to apply the LIM. 
This is
actually relevant since, even if approximate functionals are used, the first term on the right-hand side of Eq.~(\ref{eq:exact_GIC_ener})
will always be
linear in $w$. Combining LIM with the latter equation within the approximations (i) and (ii) leads to the
following ghost-interaction corrected (GIC) ensemble energy expression,
\begin{eqnarray}
\begin{array}{l}\label{eq:GIC_ens_energy}
E^{w} \rightarrow \subsupi{\tilde{E}}{GIC}{\mu,{\it w}} = 
\text{Tr}\left[ \hat{\gamma}^{\mu,{\it w}} \hat{H} \right] + 
\subsupi{E}{c,md}{sr,\mu}[n_{\hat{\gamma}^{\mu,{\it w}}}],
\end{array}
\end{eqnarray}
and to the corresponding GIC-LIM excitation energy:
\begin{eqnarray}
\begin{array}{l}\label{eq:GIC-LIM_XE}
\omega \rightarrow\tilde{\omega}^\mu_{\rm{GIC-LIM}} = 
2\left(
\subsupi{\tilde{E}}{GIC}{\mu,{\it w}=1/2} - 
\subsupi{\tilde{E}}{GIC}{\mu,{\it w}=0}  
\right).
\end{array}
\end{eqnarray}
Note that LIM (see Eqs.~(\ref{eq:widfa_ens_en}) and (\ref{eq:widfa_LIM})) and GIC-LIM excitation energies are calculated with the {\it same} (WIDFA) ensemble density matrix
$\hat{\gamma}^{\rm{\mu,{\it{w}}}}$. GIC-LIM reduces to pure wavefunction theory when
$\mu\rightarrow+\infty$. 
In the $\mu=0$ limit, the ensemble 
energy in Eq.~(\ref{eq:GIC_ens_energy}) will be
simply written as an ensemble Hartree-Fock (HF) energy (calculated with 
the KS-eDFT orbitals) complemented by the standard (full-range) density-functional correlation energy. 
In conventional ground-state DFT, the combination of $100\%$ of HF exchange with local or
semi-local correlation functionals does not work well. As discussed in
Ref.~\cite{Cornaton2013PRA}, the situation is different in the context
of ground-state range-separated DFT. Regarding excited states, in the light of the numerical
results in Sec.~\ref{sec:results_discussion}, the use of $100\%$ of HF
exchange actually improves on the accuracy of excitation energies in practical KS-eDFT
calculations. This should obviously be investigated further on more 
atomic and molecular systems.\\ 

Following Savin~\cite{savin2014towards}, we finally propose to improve
GIC-LIM further by means of 
extrapolation techniques. While the LIM excitation energy varies as $\mu^{-2}$ when
$\mu\rightarrow+\infty$~\cite{extrapol_edft}, the GIC-LIM one will vary
as $\mu^{-3}$~\cite{Toulouse2005TCA}, thus leading to the extrapolated
LIM (ELIM)~\cite{extrapol_edft} and extrapolated GIC-LIM (EGIC-LIM) excitation energy
expressions,   
\begin{eqnarray}\label{eq:extrapol_excener_2ndorder}
\tilde{\omega}^\mu_{{\rm ELIM}} &=&\tilde{\omega}^\mu_{{\rm
LIM}}+\dfrac{\mu}{2}
\dfrac{\partial \tilde{\omega}^\mu_{{\rm LIM}}}{\partial \mu},
\nonumber
\\
\tilde{\omega}^\mu_{{\rm EGIC-LIM}} &=&\tilde{\omega}^\mu_{{\rm
GIC-LIM}}+\dfrac{\mu}{3}
\dfrac{\partial \tilde{\omega}^\mu_{{\rm GIC-LIM}}}{\partial \mu}.
\end{eqnarray}
Note that GIC-LIM and EGIC-LIM schemes can
be extended to higher excitations straightforwardly by using
Eq.~(\ref{eq:generalizedWIDFA_LIM_exen}) in conjunction with 
GIC equiensemble energies.
 
\section{Computational details}\label{sec:comp_details}
All the calculations have been performed with a
development version of the DALTON program
package~\cite{DALTON,DALTON_short} on a small test set of atoms and molecules 
consisting of He, Be, H$_2$ ($R = 1.4a_{0}, 3.7a_{0})$ and HeH$^{+} (R = 8.0a_{0})$. 
The following two-state singlet ensembles in a given space symmetry have been
considered: $\{1^{1}S,2^{1}S \}$ for He
and Be, $\{ 1^{1}\Sigma^{+}, 2^{1}\Sigma^{+} \}$ for the stretched
HeH$^{+}$ molecule and $\{ 1^{1}\Sigma^{+}_{g}, 2
^{1}\Sigma^{+}_{g} \}$ for H$_2$. Note that the $1^{1}\Sigma^{+}
\rightarrow 2^{1}\Sigma^{+}$ excitation in the stretched HeH$^{+}$
molecule is a charge transfer excitation while the
$1^{1}\Sigma^{+}_{g}\rightarrow
2^{1}\Sigma^{+}_{g}$ excitation in the stretched H$_2$ molecule is a
double excitation. In order to illustrate the extension
of GIC-LIM and EGIC-LIM to higher excitations, the four-state ensemble 
$\{1^{1}S,2^{1}S,1^{1}D\}$ in $A_g$ symmetry 
has been considered in Be.
The excitation $1^{1}S \rightarrow 1^{1}D$
is a double excitation. In this case, the
ground $1^{1}S$ and first-excited $2^{1}S$ states are not degenerate ($g_{0}=1$ and
$g_{1}=1$) while the second excited state $1^{1}D$ in $A_g$ symmetry is
degenerate twice ($g_{2}=2$), thus leading to the following expression
for the $1^{1}S \rightarrow 1^{1}D$ LIM excitation energy,
according to Eq.~(\ref{eq:generalizedWIDFA_LIM_exen}),
\begin{eqnarray}\label{eq:1s1d_ex_en}
\tilde{\omega}_{{\rm{LIM}},2}^{\mu} = 
2\left(\tilde{E}_{2}^{\mu,1/4} - \tilde{E}_{1}^{\mu,1/2}\right)
+ \frac{1}{2}\tilde{\omega}_{{\rm{LIM}},1}^{\mu},
\end{eqnarray}
where $\tilde{\omega}_{{\rm{LIM}},1}^{\mu} = 
2( \tilde{E}_{1}^{\mu,1/2} - \tilde{E}_{0}^{\mu,1})$ corresponds to the
$1^{1}S\rightarrow 2^{1}S$
excitation energy.
Wavefunctions have been computed at the full configuration interaction
(FCI) level in one-electron basis sets of augmented quadruple-$\zeta$\ quality 
(aug-cc-pVQZ)~\cite{dunning1989gaussian,woon1994gaussian}. Therefore,
range-separated eDFT excitation energies will all converge
towards FCI values when increasing $\mu$.  
Long-range interacting ensemble density matrices have been computed
self-consistently at the
WIDFA level with the short-range
LDA xc potential of Savin and coworkers~\cite{savinbook,toulda}. The
corresponding xc functional was used to compute LIM excitation energies.
Finally, the ground-state md short-range correlation functional of Paziani
{\etal}~\cite{Paziani2006PRB} was used for computing GIC-LIM
excitation energies. Let us stress once more that both LIM and GIC-LIM use
exactly the same  
long-range interacting ensemble density matrix, {\it i.e.}~the one optimized at
the WIDFA level (see Eq.~(\ref{eq:widfa_ens_en})).    

\section{Results and discussion}\label{sec:results_discussion}
In Fig.~\ref{fig:XE_full_range_corr}, we have analyzed the weight-dependence of WIDFA 
auxiliary excitation energies (see Eq.~(\ref{eq:widfaXE})) and the FOC excitation energies 
(see Eq.~(\ref{eq:full-range-corrected_XE})) 
for  $\mu=0$ (KS-eDFT) and the usual $\mu=0.4a_{0}^{-1}$
value~\cite{senjean2015linear,extrapol_edft}. Although short-range correlation 
effects are neglected in FOC energies, following this 
approximation improves the accuracy of the excitation energy and reduces its weight-dependence
significantly in comparison to the WIDFA auxiliary
excitation energy.   
\begin{figure}
    \centering
\includegraphics[scale=1.0]{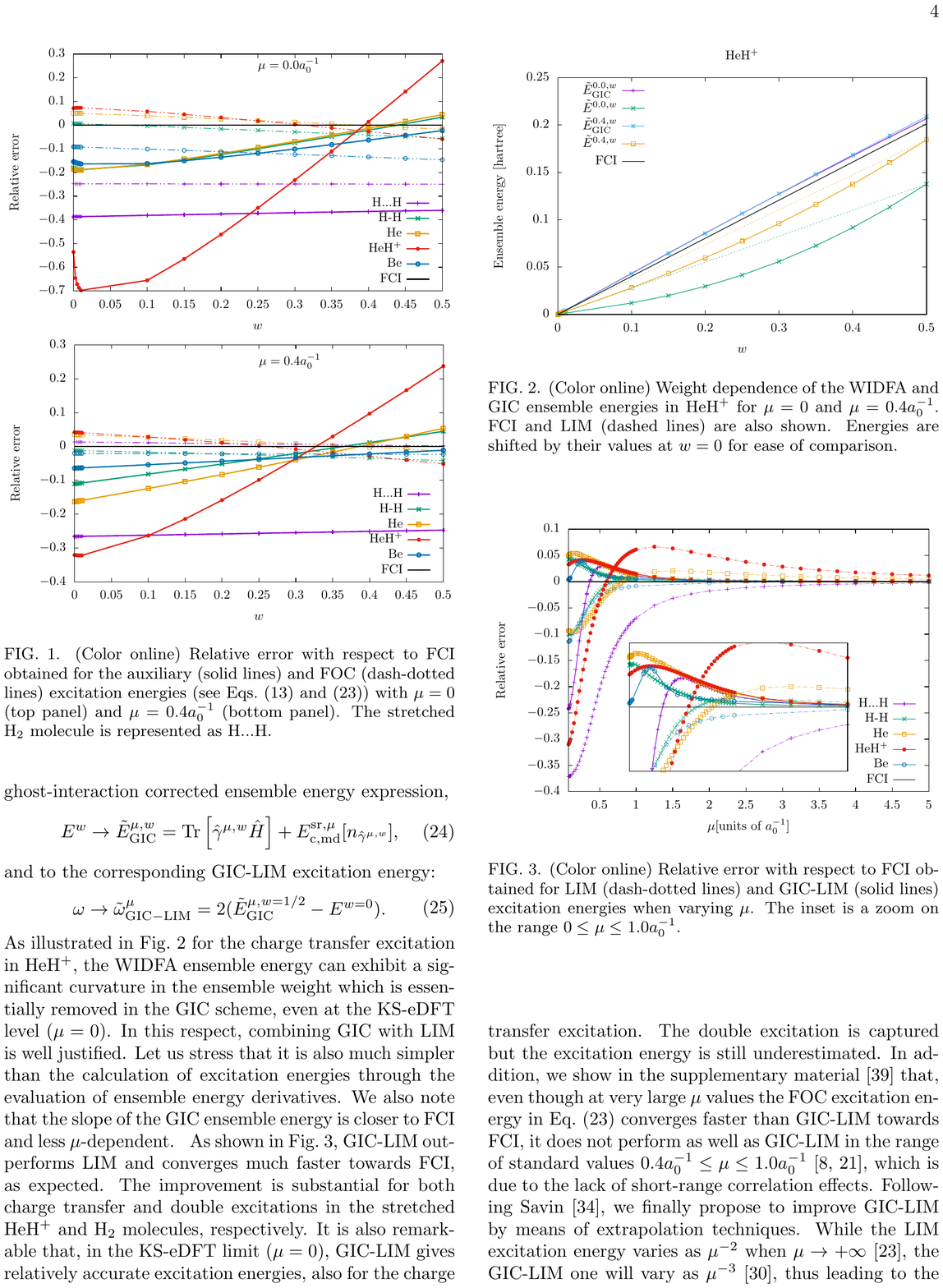}
    \caption{(Color online) Relative error with respect to FCI obtained
for the auxiliary (solid lines) and FOC (dash-dotted lines) excitation
energies 
 with $\mu = 0$ (top panel)
and $\mu = 0.4a_{0}^{-1}$ (bottom panel). The stretched H$_{2}$ molecule
is represented as H...H. Relative errors are calculated as
$\frac{\tilde{\omega} - \omega_{\rm{FCI}}}{\omega_{\rm{FCI}}}$ where
$\tilde{\omega}$ is the approximate excitation energy.
}\label{fig:XE_full_range_corr}
\end{figure}

As illustrated in Fig.~\ref{fig:curvature_hehplus} for the charge
transfer excitation $1^{1}\Sigma^{+}\rightarrow 2^{1}\Sigma^{+}$ in the
stretched HeH$^{+}$ molecule, the WIDFA ensemble energy can exhibit
a significant curvature in the ensemble weight. This is
known~\cite{senjean2015linear} and actually expected from the
expression of the ensemble short-range Hartree energy in
Eq.~(\ref{eq:gi_expression}). As expected from
Eq.~(\ref{eq:exact_GIC_ener}), the curvature is essentially
removed in the GIC scheme, even in the KS-eDFT limit ($\mu=0$). 
In this respect, combining GIC with LIM is 
well justified. Let us stress that it is also much simpler than the
calculation of excitation energies through the evaluation of ensemble energy
derivatives (see Eq.~(\ref{eq:ExEn_deri_GIC})). We also note that the slope 
of the GIC ensemble energy is closer to FCI and less $\mu$-dependent.  
\begin{figure}
    \centering
\includegraphics[scale=0.7]{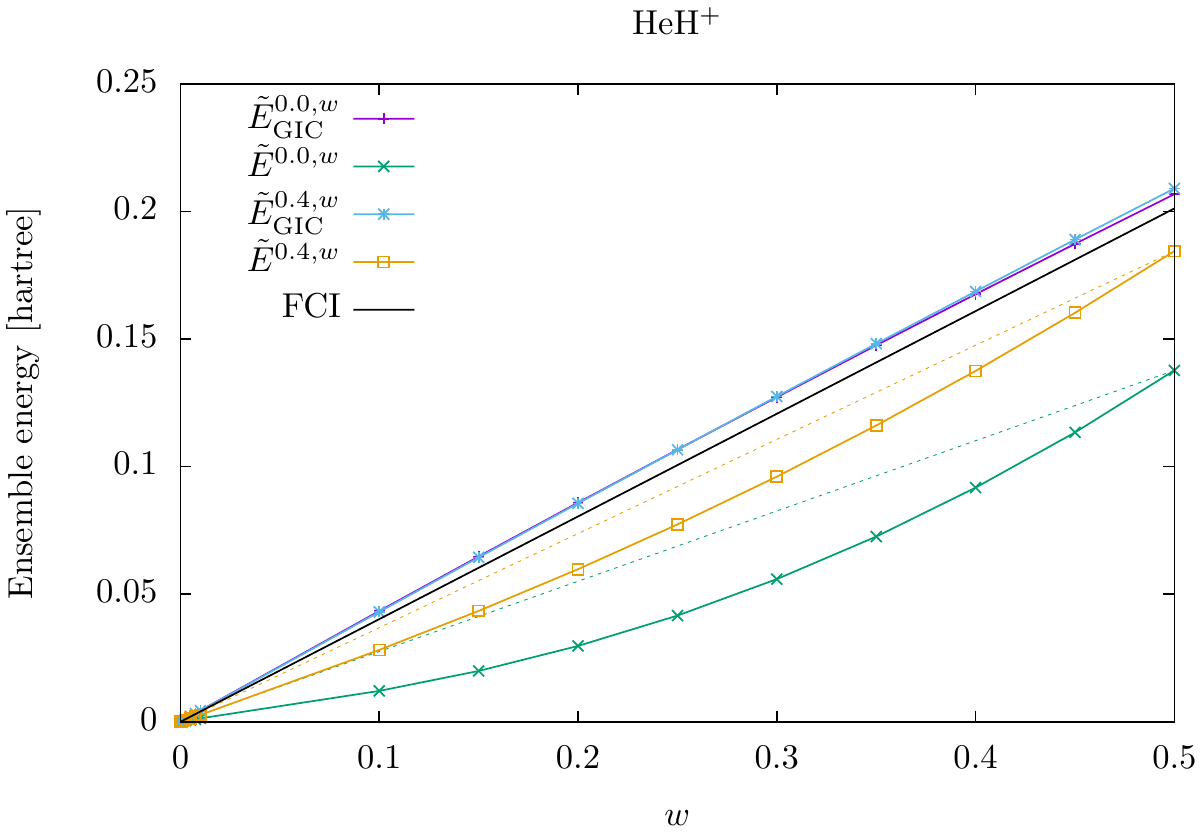}
    \caption{(Color online) Weight dependence of the WIDFA and GIC
ensemble energies in HeH$^{+}$ for $\mu=0$ and $\mu=0.4a_{0}^{-1}$. FCI
and LIM (dashed lines) are also shown. Energies are shifted by
their values at $w=0$ for ease of
comparison.}\label{fig:curvature_hehplus}
\end{figure}
\begin{figure}
    \centering
\includegraphics[scale=0.7]{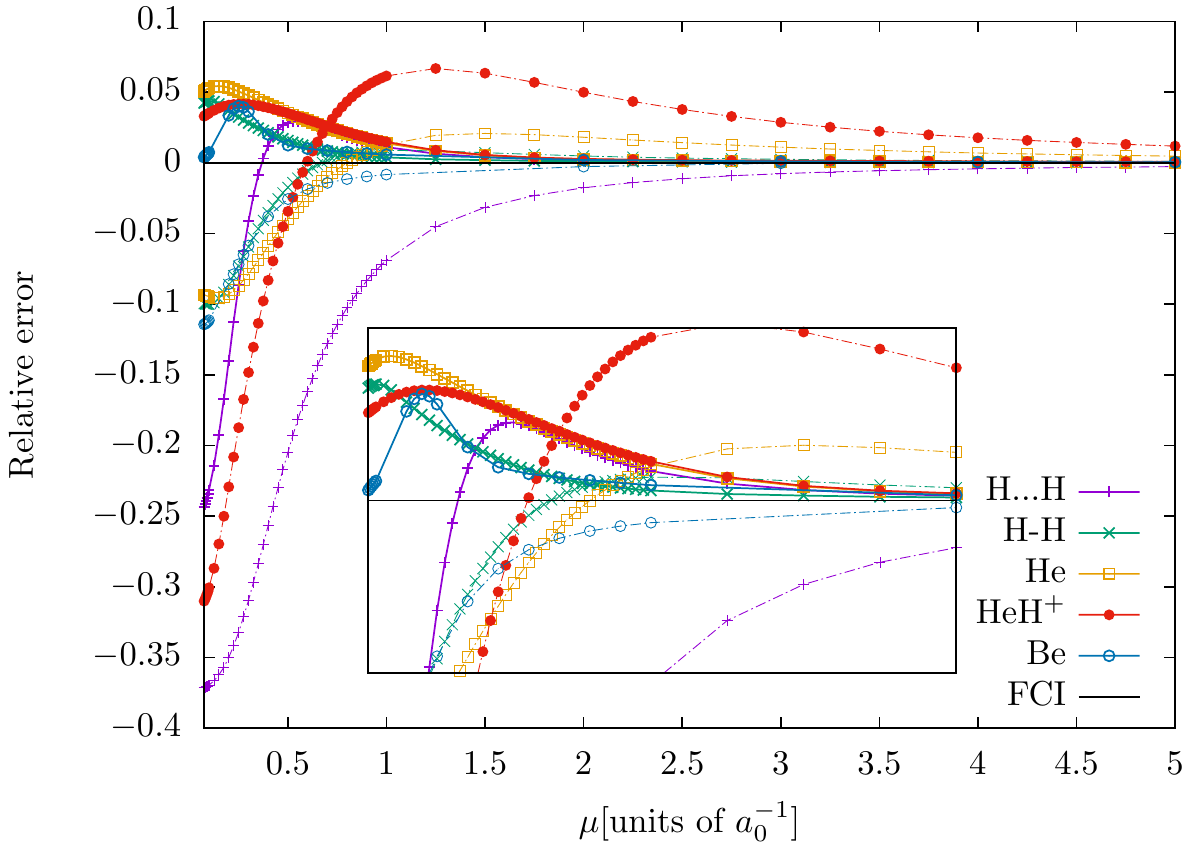}
    \caption{(Color online) Relative error with respect to FCI obtained
for LIM (dash-dotted lines) and GIC-LIM (solid lines) excitation
energies when varying $\mu$. The inset is a zoom on the range
$0\leq\mu\leq1.5a_{0}^{-1}$. Excitations in the stretched HeH$^{+}$
($1^{1}\Sigma^{+} \rightarrow 2^{1}\Sigma^{+}$) and
H$\ldots$H ($1^{1}\Sigma_g^{+} \rightarrow 2^{1}\Sigma_g^{+}$) molecules correspond to a charge transfer and a double
excitation, respectively.
}\label{fig:gic-lim_performance}
\end{figure}
\begin{figure}
    \centering
\includegraphics[scale=0.65]{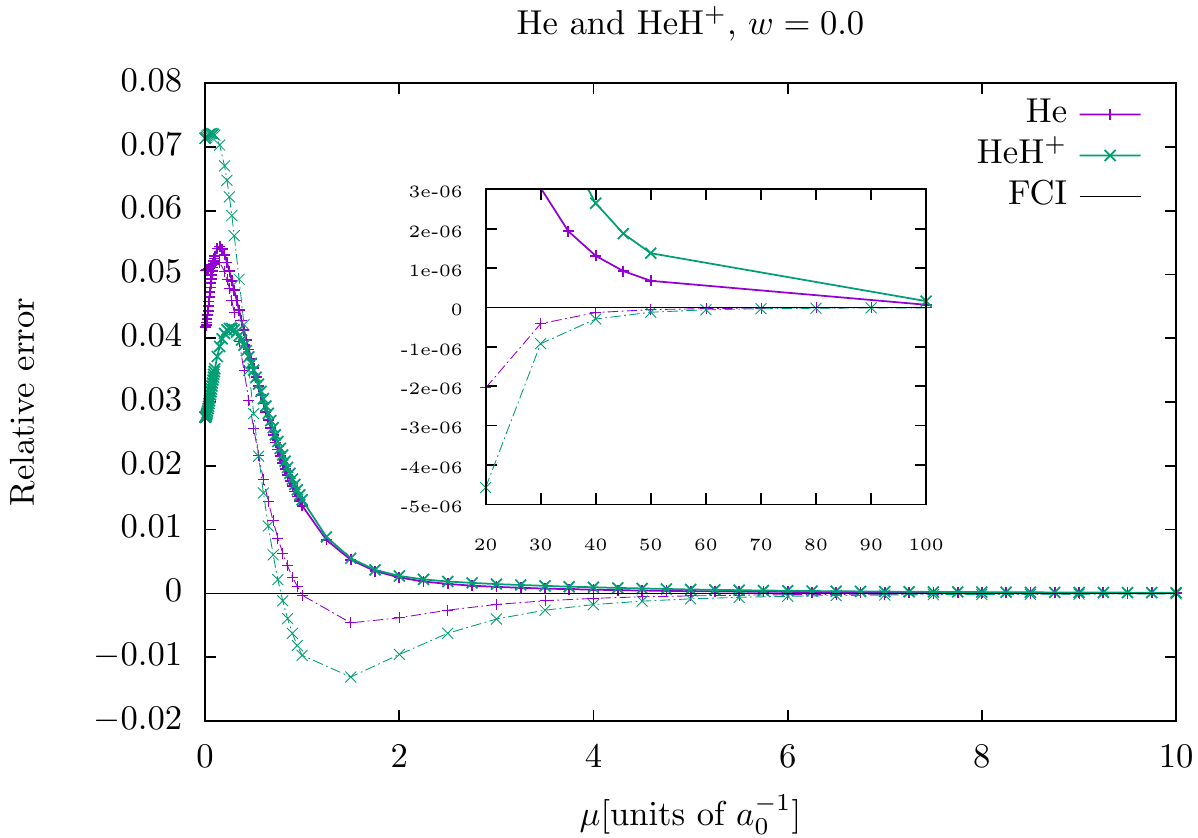}
    \caption{(Color online) 
Relative error with respect to 
    FCI obtained for GIC-LIM (solid lines) and FOC 
    [$w=0$] (dash-dotted lines) excitation energies 
    in He and the stretched HeH$^{+}$ molecule. The convergence 
    towards FCI when $\mu\rightarrow+\infty$
    is shown in the inset.
}\label{fig:conver1}
\end{figure}
\begin{figure}
    \centering
\includegraphics[scale=0.65]{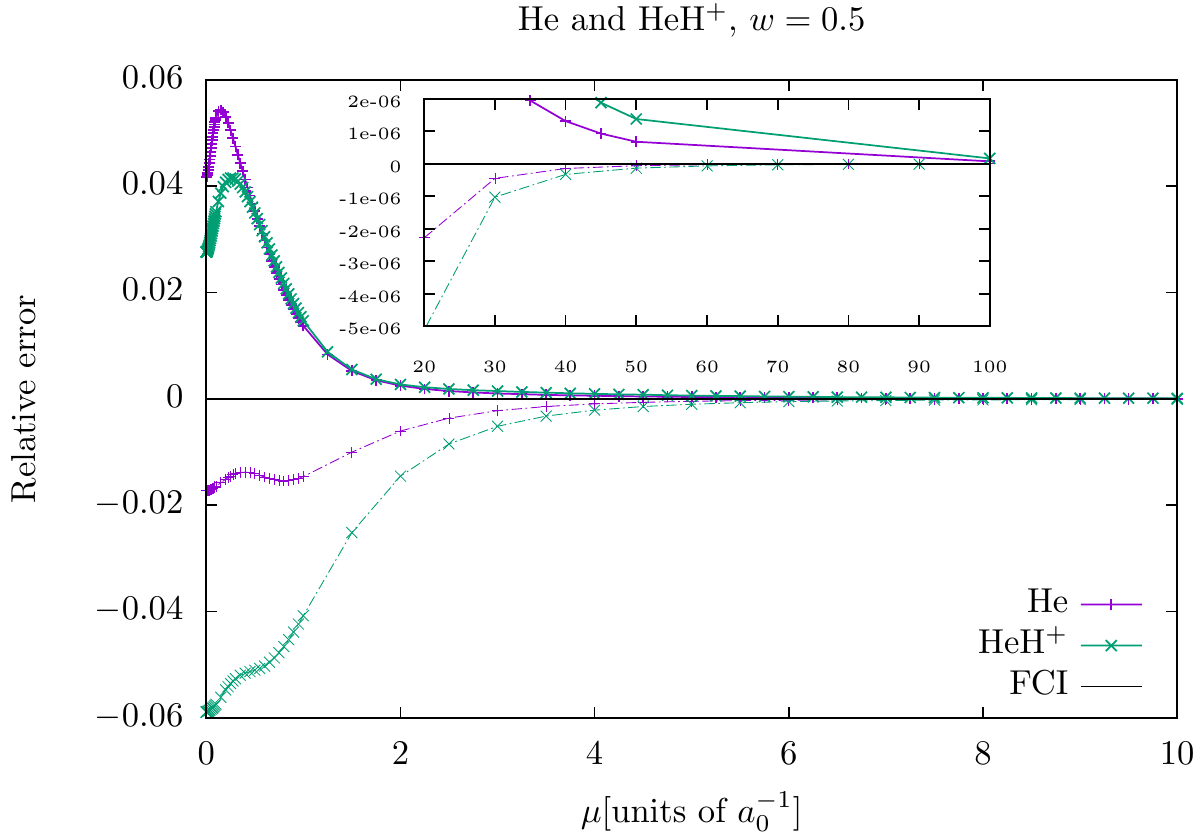}
    \caption{(Color online) 
Relative error with respect to 
    FCI obtained for GIC-LIM (solid lines) and FOC [$w=0.5$]
    (dash-dotted lines) excitation energies 
     in He and the stretched HeH$^{+}$ molecule. The convergence 
    towards FCI when $\mu\rightarrow+\infty$
    is shown in the inset.
}\label{fig:conver2}
\end{figure}
\\
As shown in Fig.~\ref{fig:gic-lim_performance}, GIC-LIM outperforms LIM and
converges much faster towards FCI when increasing $\mu$, as expected. The
improvement is substantial for both charge transfer and double
excitations in the stretched HeH$^{+}$ and H$_2$ molecules,
respectively. It is also remarkable that, in the KS-eDFT limit
($\mu=0$), GIC-LIM gives relatively accurate excitation energies, also
for the charge transfer excitation, and despite the fact that 100$\%$ of
HF exchange is combined with an LDA correlation functional. The double excitation in H...H is captured
but the excitation energy is still underestimated. In addition, as shown in 
Figs.~\ref{fig:conver1} and \ref{fig:conver2} for $w=0$ and $w=0.5$,
respectively, even though at very
 large $\mu$ values the FOC excitation energy converges faster 
 than GIC-LIM towards FCI, it does not necessarily perform better than 
 GIC-LIM in the range of standard values $0.4a_{0}^{-1}\leq\mu\leq1.0a_{0}^{-1}$
 ~\cite{PRA13_Pernal_srEDFT,senjean2015linear}, 
which is due to the lack of short-range correlation effects. 
\begin{figure}
    \centering
\includegraphics[scale=0.7]{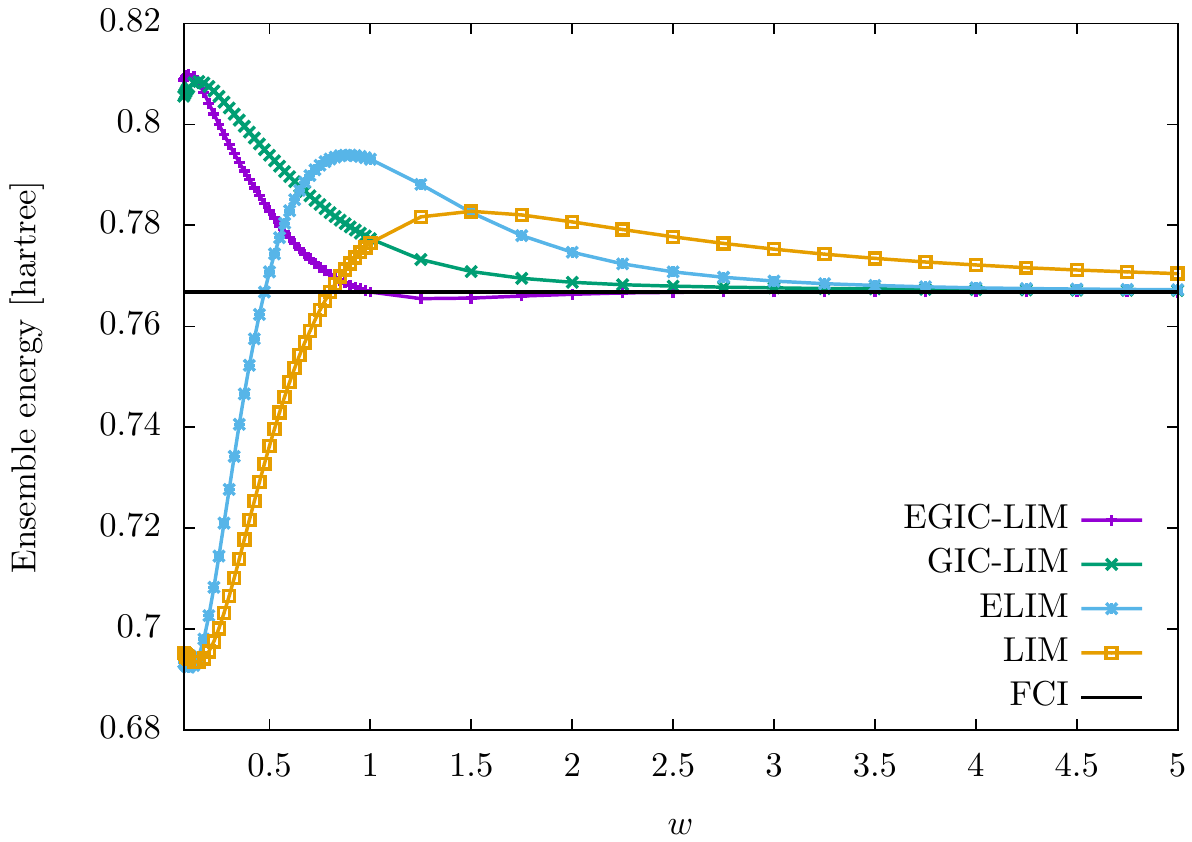}
    \caption{(Color online) LIM and GIC-LIM $1^{1}S\rightarrow 2^{1}S$ excitation
energies obtained in He with and without extrapolation corrections when
varying $\mu$. Comparison is made with FCI. See text for further details.}\label{fig:EGIC-LIM}
\end{figure}
\begin{figure}
    \centering
\includegraphics[scale=1.1]{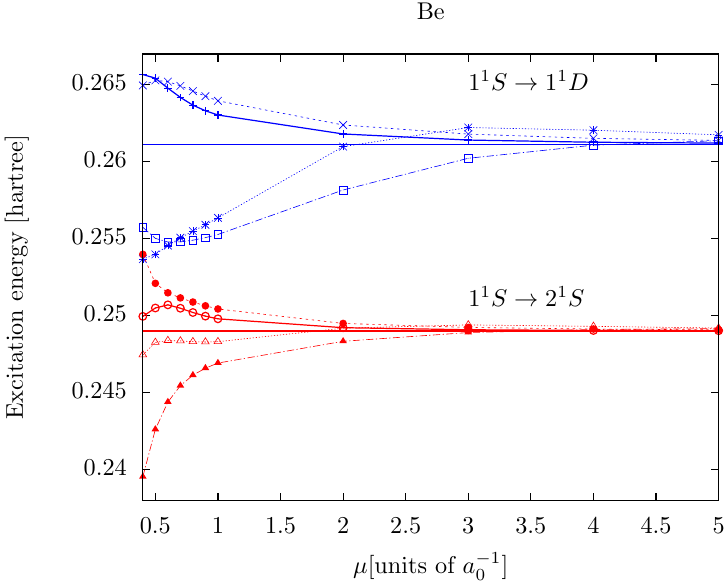}
    \caption{(Color online) Convergence towards FCI (solid unmarked
straight lines) of the LIM (marked dash-dotted lines), 
    ELIM (marked dotted lines), GIC-LIM (marked dashed lines) and 
    EGIC-LIM (marked solid lines) excitation energies obtained for 
    the singly-excited $1^{1}S \rightarrow 2^{1}S$ (bottom red curves) and
doubly-excited $1^{1}S \rightarrow 1^{1}D$ (top blue curves) 
    transitions in Be when increasing $\mu$.}\label{fig:be_1d}
\end{figure}

Finally, the effect of extrapolation on the GIC-LIM  excitation energy 
is shown for He in Fig.~\ref{fig:EGIC-LIM}. When increasing $\mu$ from
$0.2a_0^{-1}$, EGIC-LIM converges monotonically towards FCI very
rapidly, in contrast to GIC-LIM and even ELIM. Convergence is almost reached at the standard $\mu=1.0a_0^{-1}$
value~\cite{PRA13_Pernal_srEDFT}. It means that accurate
ghost-interaction free excitation
energies can in principle be obtained with a relatively small $\mu$ value
which is highly desirable. Indeed, if $\mu$ is not too large, ground-
and excited-state long-range interacting wavefunctions are expected to have a
rather compact configuration expansion. Convergence with respect to the
atomic basis set will also be
faster~\cite{JCP15_Odile_basis_convergence_srDFT}.
In order to illustrate the extension of (E)GIC-LIM to higher
excitations, we consider the double excitation 
$1^{1}S \rightarrow 1^{1}D$ 
in Be. Results are shown in Fig.~\ref{fig:be_1d}. 
We see that the convergence towards FCI of EGIC-LIM is slightly slower for the double excitation  
than for the single $1^{1}S \rightarrow 2^{1}S$ excitation. 
Nevertheless, results are still accurate for both excitations  
in the range of standard values $0.4a_{0}^{-1}\leq\mu\leq1.0a_{0}^{-1}$.

\section{Conclusions}\label{sec:conclu}

A rigorous ghost interaction correction (GIC) scheme has been proposed
in the context of range-separated ensemble 
density-functional theory (eDFT). It is based on an exact decomposition
of the short-range ensemble exchange-correlation energy into a
multideterminantal exact exchange contribution and a complementary
density-functional correlation energy for which an adiabatic 
connection formula has been derived. 
In order to perform practical calculations, the
latter correlation functional has been simply modeled by its ground-state
local density approximation (LDA) while the long-range interacting ensemble
density matrix is obtained self-consistently by combining a long-range
configuration interaction calculation with a short-range LDA potential.    
Excitation energies can then be computed from the GIC ensemble energies
by means of a linear interpolation method (LIM) with, on top, an extrapolation
correction. 
Results have been shown for He, Be and small molecular systems 
(H$_2$ and HeH$^+$). While providing approximate ensemble energies that are
essentially linear in the ensemble weight, the GIC scheme 
gives a significant improvement on the accuracy of excitation energies.  
In particular, the charge-transfer excitation 
$1^{1}\Sigma^{+}\rightarrow 2^{1}\Sigma^{+}$ in the stretched HeH$^+$
molecule as well as 
the double excitation $1^{1}S\rightarrow 1^{1}D$ in Be are well
described for standard range-separation parameter values. Interestingly,
relatively good results are also obtained when the latter parameter is
set to zero, which corresponds to standard Kohn--Sham (KS) eDFT. In this
case, the GIC ensemble energy reduces to an ensemble Hartree--Fock
energy (calculated with the ensemble KS orbitals) complemented by a
local density-functional correlation energy. Test calculations on larger
systems shoud be performed in order to assess the reliability of the   
GIC approach, in particular in fields like photochemistry where the use
of ensembles and range separation is appealing. It would also be
interesting to construct weight-dependent correlation functionals
along the proposed generalized adiabatic connection for ensembles and to
remove from our current GIC scheme the residual ghost-correlation error.
Work is currently in progress in these directions. 

\section{Acknowledgements}
M. M. Alam acknowledges Bruno Senjean for fruitful discussions and help. 
The authors acknowledge financial support from the LABEX `Chemistry 
of complex systems' and the ANR (MCFUNEX project).

\begin{thebibliography}{43}%
\makeatletter
\providecommand \@ifxundefined [1]{%
 \@ifx{#1\undefined}
}%
\providecommand \@ifnum [1]{%
 \ifnum #1\expandafter \@firstoftwo
 \else \expandafter \@secondoftwo
 \fi
}%
\providecommand \@ifx [1]{%
 \ifx #1\expandafter \@firstoftwo
 \else \expandafter \@secondoftwo
 \fi
}%
\providecommand \natexlab [1]{#1}%
\providecommand \enquote  [1]{``#1''}%
\providecommand \bibnamefont  [1]{#1}%
\providecommand \bibfnamefont [1]{#1}%
\providecommand \citenamefont [1]{#1}%
\providecommand \href@noop [0]{\@secondoftwo}%
\providecommand \href [0]{\begingroup \@sanitize@url \@href}%
\providecommand \@href[1]{\@@startlink{#1}\@@href}%
\providecommand \@@href[1]{\endgroup#1\@@endlink}%
\providecommand \@sanitize@url [0]{\catcode `\\12\catcode `\$12\catcode
  `\&12\catcode `\#12\catcode `\^12\catcode `\_12\catcode `\%12\relax}%
\providecommand \@@startlink[1]{}%
\providecommand \@@endlink[0]{}%
\providecommand \url  [0]{\begingroup\@sanitize@url \@url }%
\providecommand \@url [1]{\endgroup\@href {#1}{\urlprefix }}%
\providecommand \urlprefix  [0]{URL }%
\providecommand \Eprint [0]{\href }%
\providecommand \doibase [0]{http://dx.doi.org/}%
\providecommand \selectlanguage [0]{\@gobble}%
\providecommand \bibinfo  [0]{\@secondoftwo}%
\providecommand \bibfield  [0]{\@secondoftwo}%
\providecommand \translation [1]{[#1]}%
\providecommand \BibitemOpen [0]{}%
\providecommand \bibitemStop [0]{}%
\providecommand \bibitemNoStop [0]{.\EOS\space}%
\providecommand \EOS [0]{\spacefactor3000\relax}%
\providecommand \BibitemShut  [1]{\csname bibitem#1\endcsname}%
\let\auto@bib@innerbib\@empty
\bibitem [{\citenamefont {Casida}\ and\ \citenamefont
  {Huix-Rotllant}(2012)}]{Casida_tddft_review_2012}%
  \BibitemOpen
  \bibfield  {author} {\bibinfo {author} {\bibfnamefont {M.}~\bibnamefont
  {Casida}}\ and\ \bibinfo {author} {\bibfnamefont {M.}~\bibnamefont
  {Huix-Rotllant}},\ }\href@noop {} {\bibfield  {journal} {\bibinfo  {journal}
  {Annu. Rev. Phys. Chem.}\ }\textbf {\bibinfo {volume} {63}},\ \bibinfo
  {pages} {287} (\bibinfo {year} {2012})}\BibitemShut {NoStop}%
\bibitem [{\citenamefont {Marques}\ and\ \citenamefont
  {Gross}(2004)}]{marques2004time}%
  \BibitemOpen
  \bibfield  {author} {\bibinfo {author} {\bibfnamefont {M.}~\bibnamefont
  {Marques}}\ and\ \bibinfo {author} {\bibfnamefont {E.}~\bibnamefont
  {Gross}},\ }\href@noop {} {\bibfield  {journal} {\bibinfo  {journal} {Annu.
  Rev. Phys. Chem.}\ }\textbf {\bibinfo {volume} {55}},\ \bibinfo {pages} {427}
  (\bibinfo {year} {2004})}\BibitemShut {NoStop}%
\bibitem [{\citenamefont {Maitra}\ \emph {et~al.}(2004)\citenamefont {Maitra},
  \citenamefont {Zhang}, \citenamefont {Cave},\ and\ \citenamefont
  {Burke}}]{maitra2004double}%
  \BibitemOpen
  \bibfield  {author} {\bibinfo {author} {\bibfnamefont {N.~T.}\ \bibnamefont
  {Maitra}}, \bibinfo {author} {\bibfnamefont {F.}~\bibnamefont {Zhang}},
  \bibinfo {author} {\bibfnamefont {R.~J.}\ \bibnamefont {Cave}}, \ and\
  \bibinfo {author} {\bibfnamefont {K.}~\bibnamefont {Burke}},\ }\href@noop {}
  {\bibfield  {journal} {\bibinfo  {journal} {J. Chem. Phys.}\ }\textbf
  {\bibinfo {volume} {120}},\ \bibinfo {pages} {5932} (\bibinfo {year}
  {2004})}\BibitemShut {NoStop}%
\bibitem [{\citenamefont {Theophilou}(1979)}]{JPC79_Theophilou_equi-ensembles}%
  \BibitemOpen
  \bibfield  {author} {\bibinfo {author} {\bibfnamefont {A.~K.}\ \bibnamefont
  {Theophilou}},\ }\href@noop {} {\bibfield  {journal} {\bibinfo  {journal} {J.
  Phys. C (Solid State Phys.)}\ }\textbf {\bibinfo {volume} {12}},\ \bibinfo
  {pages} {5419} (\bibinfo {year} {1979})}\BibitemShut {NoStop}%
\bibitem [{\citenamefont {Gross}\ \emph
  {et~al.}(1988{\natexlab{a}})\citenamefont {Gross}, \citenamefont {Oliveira},\
  and\ \citenamefont {Kohn}}]{PRA_GOK_RRprinc}%
  \BibitemOpen
  \bibfield  {author} {\bibinfo {author} {\bibfnamefont {E.~K.~U.}\
  \bibnamefont {Gross}}, \bibinfo {author} {\bibfnamefont {L.~N.}\ \bibnamefont
  {Oliveira}}, \ and\ \bibinfo {author} {\bibfnamefont {W.}~\bibnamefont
  {Kohn}},\ }\href {\doibase 10.1103/PhysRevA.37.2805} {\bibfield  {journal}
  {\bibinfo  {journal} {Phys. Rev. A}\ }\textbf {\bibinfo {volume} {37}},\
  \bibinfo {pages} {2805} (\bibinfo {year} {1988}{\natexlab{a}})}\BibitemShut
  {NoStop}%
\bibitem [{\citenamefont {Gross}\ \emph
  {et~al.}(1988{\natexlab{b}})\citenamefont {Gross}, \citenamefont {Oliveira},\
  and\ \citenamefont {Kohn}}]{PRA_GOK_EKSDFT}%
  \BibitemOpen
  \bibfield  {author} {\bibinfo {author} {\bibfnamefont {E.~K.~U.}\
  \bibnamefont {Gross}}, \bibinfo {author} {\bibfnamefont {L.~N.}\ \bibnamefont
  {Oliveira}}, \ and\ \bibinfo {author} {\bibfnamefont {W.}~\bibnamefont
  {Kohn}},\ }\href {\doibase 10.1103/PhysRevA.37.2809} {\bibfield  {journal}
  {\bibinfo  {journal} {Phys. Rev. A}\ }\textbf {\bibinfo {volume} {37}},\
  \bibinfo {pages} {2809} (\bibinfo {year} {1988}{\natexlab{b}})}\BibitemShut
  {NoStop}%
\bibitem [{\citenamefont {Gross}\ \emph
  {et~al.}(1988{\natexlab{c}})\citenamefont {Gross}, \citenamefont {Oliveira},\
  and\ \citenamefont {Kohn}}]{GOK3}%
  \BibitemOpen
  \bibfield  {author} {\bibinfo {author} {\bibfnamefont {E.~K.~U.}\
  \bibnamefont {Gross}}, \bibinfo {author} {\bibfnamefont {L.~N.}\ \bibnamefont
  {Oliveira}}, \ and\ \bibinfo {author} {\bibfnamefont {W.}~\bibnamefont
  {Kohn}},\ }\href {\doibase 10.1103/PhysRevA.37.2821} {\bibfield  {journal}
  {\bibinfo  {journal} {Phys. Rev. A}\ }\textbf {\bibinfo {volume} {37}},\
  \bibinfo {pages} {2821} (\bibinfo {year} {1988}{\natexlab{c}})}\BibitemShut
  {NoStop}%
\bibitem [{\citenamefont {Pastorczak}\ \emph {et~al.}(2013)\citenamefont
  {Pastorczak}, \citenamefont {Gidopoulos},\ and\ \citenamefont
  {Pernal}}]{PRA13_Pernal_srEDFT}%
  \BibitemOpen
  \bibfield  {author} {\bibinfo {author} {\bibfnamefont {E.}~\bibnamefont
  {Pastorczak}}, \bibinfo {author} {\bibfnamefont {N.~I.}\ \bibnamefont
  {Gidopoulos}}, \ and\ \bibinfo {author} {\bibfnamefont {K.}~\bibnamefont
  {Pernal}},\ }\href@noop {} {\bibfield  {journal} {\bibinfo  {journal} {Phys.
  Rev. A}\ }\textbf {\bibinfo {volume} {87}},\ \bibinfo {pages} {062501}
  (\bibinfo {year} {2013})}\BibitemShut {NoStop}%
\bibitem [{\citenamefont {Franck}\ and\ \citenamefont
  {Fromager}(2014)}]{ac_fromager}%
  \BibitemOpen
  \bibfield  {author} {\bibinfo {author} {\bibfnamefont {O.}~\bibnamefont
  {Franck}}\ and\ \bibinfo {author} {\bibfnamefont {E.}~\bibnamefont
  {Fromager}},\ }\href@noop {} {\bibfield  {journal} {\bibinfo  {journal} {Mol.
  Phys.}\ }\textbf {\bibinfo {volume} {112}},\ \bibinfo {pages} {1684}
  (\bibinfo {year} {2014})}\BibitemShut {NoStop}%
\bibitem [{\citenamefont {Pribram-Jones}\ \emph {et~al.}(2014)\citenamefont
  {Pribram-Jones}, \citenamefont {hui Yang}, \citenamefont {R.Trail},
  \citenamefont {Burke}, \citenamefont {J.Needs},\ and\ \citenamefont
  {A.Ullrich}}]{Burke_ensemble}%
  \BibitemOpen
  \bibfield  {author} {\bibinfo {author} {\bibfnamefont {A.}~\bibnamefont
  {Pribram-Jones}}, \bibinfo {author} {\bibfnamefont {Z.}~\bibnamefont {hui
  Yang}}, \bibinfo {author} {\bibfnamefont {J.}~\bibnamefont {R.Trail}},
  \bibinfo {author} {\bibfnamefont {K.}~\bibnamefont {Burke}}, \bibinfo
  {author} {\bibfnamefont {R.}~\bibnamefont {J.Needs}}, \ and\ \bibinfo
  {author} {\bibfnamefont {C.}~\bibnamefont {A.Ullrich}},\ }\href {\doibase
  10.1063/1.4872255} {\bibfield  {journal} {\bibinfo  {journal} {J. Chem.
  Phys.}\ }\textbf {\bibinfo {volume} {140}},\ \bibinfo {pages} {18A541}
  (\bibinfo {year} {2014})}\BibitemShut {NoStop}%
\bibitem [{\citenamefont {Yang}\ \emph {et~al.}(2014)\citenamefont {Yang},
  \citenamefont {Trail}, \citenamefont {Pribram-Jones}, \citenamefont {Burke},
  \citenamefont {Needs},\ and\ \citenamefont {Ullrich}}]{yang2014exact}%
  \BibitemOpen
  \bibfield  {author} {\bibinfo {author} {\bibfnamefont {Z.-h.}\ \bibnamefont
  {Yang}}, \bibinfo {author} {\bibfnamefont {J.~R.}\ \bibnamefont {Trail}},
  \bibinfo {author} {\bibfnamefont {A.}~\bibnamefont {Pribram-Jones}}, \bibinfo
  {author} {\bibfnamefont {K.}~\bibnamefont {Burke}}, \bibinfo {author}
  {\bibfnamefont {R.~J.}\ \bibnamefont {Needs}}, \ and\ \bibinfo {author}
  {\bibfnamefont {C.~A.}\ \bibnamefont {Ullrich}},\ }\href@noop {} {\bibfield
  {journal} {\bibinfo  {journal} {Phys. Rev. A}\ }\textbf {\bibinfo {volume}
  {90}},\ \bibinfo {pages} {042501} (\bibinfo {year} {2014})}\BibitemShut
  {NoStop}%
\bibitem [{\citenamefont {Nikiforov}\ \emph {et~al.}(2014)\citenamefont
  {Nikiforov}, \citenamefont {Gamez}, \citenamefont {Thiel}, \citenamefont
  {Huix-Rotllant},\ and\ \citenamefont
  {Filatov}}]{JCP14_Filatov_conical_inter_REKS}%
  \BibitemOpen
  \bibfield  {author} {\bibinfo {author} {\bibfnamefont {A.}~\bibnamefont
  {Nikiforov}}, \bibinfo {author} {\bibfnamefont {J.~A.}\ \bibnamefont
  {Gamez}}, \bibinfo {author} {\bibfnamefont {W.}~\bibnamefont {Thiel}},
  \bibinfo {author} {\bibfnamefont {M.}~\bibnamefont {Huix-Rotllant}}, \ and\
  \bibinfo {author} {\bibfnamefont {M.}~\bibnamefont {Filatov}},\ }\href
  {\doibase http://dx.doi.org/10.1063/1.4896372} {\bibfield  {journal}
  {\bibinfo  {journal} {J. Chem. Phys.}\ }\textbf {\bibinfo {volume} {141}},\
  \bibinfo {pages} {124122} (\bibinfo {year} {2014})}\BibitemShut {NoStop}%
\bibitem [{\citenamefont {Pastorczak}\ and\ \citenamefont
  {Pernal}(2014)}]{pastorczak2014ensemble}%
  \BibitemOpen
  \bibfield  {author} {\bibinfo {author} {\bibfnamefont {E.}~\bibnamefont
  {Pastorczak}}\ and\ \bibinfo {author} {\bibfnamefont {K.}~\bibnamefont
  {Pernal}},\ }\href@noop {} {\bibfield  {journal} {\bibinfo  {journal} {J.
  Chem. Phys.}\ }\textbf {\bibinfo {volume} {140}},\ \bibinfo {pages} {18A514}
  (\bibinfo {year} {2014})}\BibitemShut {NoStop}%
\bibitem [{\citenamefont {Filatov}\ \emph {et~al.}(2015)\citenamefont
  {Filatov}, \citenamefont {Huix-Rotllant},\ and\ \citenamefont
  {Burghardt}}]{filatov2015ensemble}%
  \BibitemOpen
  \bibfield  {author} {\bibinfo {author} {\bibfnamefont {M.}~\bibnamefont
  {Filatov}}, \bibinfo {author} {\bibfnamefont {M.}~\bibnamefont
  {Huix-Rotllant}}, \ and\ \bibinfo {author} {\bibfnamefont {I.}~\bibnamefont
  {Burghardt}},\ }\href@noop {} {\bibfield  {journal} {\bibinfo  {journal} {J.
  Chem. Phys.}\ }\textbf {\bibinfo {volume} {142}},\ \bibinfo {pages} {184104}
  (\bibinfo {year} {2015})}\BibitemShut {NoStop}%
\bibitem [{\citenamefont {Filatov}(2015)}]{Filatov-2015-Wiley}%
  \BibitemOpen
  \bibfield  {author} {\bibinfo {author} {\bibfnamefont {M.}~\bibnamefont
  {Filatov}},\ }\href {\doibase 10.1002/wcms.1209} {\bibfield  {journal}
  {\bibinfo  {journal} {WIREs Comput Mol Sci}\ }\textbf {\bibinfo {volume}
  {5}},\ \bibinfo {pages} {146} (\bibinfo {year} {2015})}\BibitemShut {NoStop}%
\bibitem [{\citenamefont {Pastorczak}\ and\ \citenamefont
  {Pernal}(2016)}]{pernal2016ghost}%
  \BibitemOpen
  \bibfield  {author} {\bibinfo {author} {\bibfnamefont {E.}~\bibnamefont
  {Pastorczak}}\ and\ \bibinfo {author} {\bibfnamefont {K.}~\bibnamefont
  {Pernal}},\ }\href {\doibase http://dx.doi.org/10.1002/qua.25107} {\bibfield
  {journal} {\bibinfo  {journal} {Int. J. Quantum Chem.}\ } (\bibinfo {year}
  {2016}),\ http://dx.doi.org/10.1002/qua.25107}\BibitemShut {NoStop}%
\bibitem [{\citenamefont {Nagy}(1996)}]{Nagy_enseXpot}%
  \BibitemOpen
  \bibfield  {author} {\bibinfo {author} {\bibfnamefont {A.}~\bibnamefont
  {Nagy}},\ }\href@noop {} {\bibfield  {journal} {\bibinfo  {journal} {J. Phys.
  B: At. Mol. Opt. Phys.}\ }\textbf {\bibinfo {volume} {29}},\ \bibinfo {pages}
  {389} (\bibinfo {year} {1996})}\BibitemShut {NoStop}%
\bibitem [{\citenamefont {Paragi}\ \emph {et~al.}(2000)\citenamefont {Paragi},
  \citenamefont {Gy\'{e}m\'{a}nt},\ and\ \citenamefont {Doren}}]{ParagiXens}%
  \BibitemOpen
  \bibfield  {author} {\bibinfo {author} {\bibfnamefont {G.}~\bibnamefont
  {Paragi}}, \bibinfo {author} {\bibfnamefont {I.}~\bibnamefont
  {Gy\'{e}m\'{a}nt}}, \ and\ \bibinfo {author} {\bibfnamefont {V.~V.}\
  \bibnamefont {Doren}},\ }\href {\doibase 10.1016/S0009-2614(00)00613-8}
  {\bibfield  {journal} {\bibinfo  {journal} {Chem. Phys. Lett.}\ }\textbf
  {\bibinfo {volume} {324}},\ \bibinfo {pages} {440} (\bibinfo {year}
  {2000})}\BibitemShut {NoStop}%
\bibitem [{\citenamefont {Paragi}\ \emph {et~al.}(2001)\citenamefont {Paragi},
  \citenamefont {Gy\'{e}m\'{a}nt},\ and\ \citenamefont {Doren}}]{ParagiXCens}%
  \BibitemOpen
  \bibfield  {author} {\bibinfo {author} {\bibfnamefont {G.}~\bibnamefont
  {Paragi}}, \bibinfo {author} {\bibfnamefont {I.}~\bibnamefont
  {Gy\'{e}m\'{a}nt}}, \ and\ \bibinfo {author} {\bibfnamefont {V.~V.}\
  \bibnamefont {Doren}},\ }\href {\doibase 10.1016/S0166-1280(01)00561-9}
  {\bibfield  {journal} {\bibinfo  {journal} {J. Mol. Struct. (Theochem)}\
  }\textbf {\bibinfo {volume} {571}},\ \bibinfo {pages} {153} (\bibinfo {year}
  {2001})}\BibitemShut {NoStop}%
\bibitem [{\citenamefont {Pernal}\ \emph {et~al.}(2015)\citenamefont {Pernal},
  \citenamefont {Gidopoulos},\ and\ \citenamefont
  {Pastorczak}}]{pernal2015excitation}%
  \BibitemOpen
  \bibfield  {author} {\bibinfo {author} {\bibfnamefont {K.}~\bibnamefont
  {Pernal}}, \bibinfo {author} {\bibfnamefont {N.~I.}\ \bibnamefont
  {Gidopoulos}}, \ and\ \bibinfo {author} {\bibfnamefont {E.}~\bibnamefont
  {Pastorczak}},\ }\href {\doibase
  http://dx.doi.org/10.1016/bs.aiq.2015.06.001} {\bibfield  {journal} {\bibinfo
   {journal} {Adv. Quantum Chem.}\ }\textbf {\bibinfo {volume} {73}},\ \bibinfo
  {pages} {199} (\bibinfo {year} {2015})}\BibitemShut {NoStop}%
\bibitem [{\citenamefont {Senjean}\ \emph {et~al.}(2015)\citenamefont
  {Senjean}, \citenamefont {Knecht}, \citenamefont {Jensen},\ and\
  \citenamefont {Fromager}}]{senjean2015linear}%
  \BibitemOpen
  \bibfield  {author} {\bibinfo {author} {\bibfnamefont {B.}~\bibnamefont
  {Senjean}}, \bibinfo {author} {\bibfnamefont {S.}~\bibnamefont {Knecht}},
  \bibinfo {author} {\bibfnamefont {H.~J.~{\Aa}.}\ \bibnamefont {Jensen}}, \
  and\ \bibinfo {author} {\bibfnamefont {E.}~\bibnamefont {Fromager}},\
  }\href@noop {} {\bibfield  {journal} {\bibinfo  {journal} {Phys. Rev. A}\
  }\textbf {\bibinfo {volume} {92}},\ \bibinfo {pages} {012518} (\bibinfo
  {year} {2015})}\BibitemShut {NoStop}%
\bibitem [{\citenamefont {Gidopoulos}\ \emph {et~al.}(2002)\citenamefont
  {Gidopoulos}, \citenamefont {Papaconstantinou},\ and\ \citenamefont
  {Gross}}]{ensemble_ghost_interaction}%
  \BibitemOpen
  \bibfield  {author} {\bibinfo {author} {\bibfnamefont {N.~I.}\ \bibnamefont
  {Gidopoulos}}, \bibinfo {author} {\bibfnamefont {P.~G.}\ \bibnamefont
  {Papaconstantinou}}, \ and\ \bibinfo {author} {\bibfnamefont {E.~K.~U.}\
  \bibnamefont {Gross}},\ }\href@noop {} {\bibfield  {journal} {\bibinfo
  {journal} {Phys. Rev. Lett.}\ }\textbf {\bibinfo {volume} {88}},\ \bibinfo
  {pages} {033003} (\bibinfo {year} {2002})}\BibitemShut {NoStop}%
\bibitem [{\citenamefont {Senjean}\ \emph {et~al.}(2016)\citenamefont
  {Senjean}, \citenamefont {Hedeg{\aa}rd}, \citenamefont {Alam}, \citenamefont
  {Knecht},\ and\ \citenamefont {Fromager}}]{extrapol_edft}%
  \BibitemOpen
  \bibfield  {author} {\bibinfo {author} {\bibfnamefont {B.}~\bibnamefont
  {Senjean}}, \bibinfo {author} {\bibfnamefont {E.~D.}\ \bibnamefont
  {Hedeg{\aa}rd}}, \bibinfo {author} {\bibfnamefont {M.~M.}\ \bibnamefont
  {Alam}}, \bibinfo {author} {\bibfnamefont {S.}~\bibnamefont {Knecht}}, \ and\
  \bibinfo {author} {\bibfnamefont {E.}~\bibnamefont {Fromager}},\ }\href@noop
  {} {\bibfield  {journal} {\bibinfo  {journal} {Mol. Phys.}\ } (\bibinfo
  {year} {2016})},\ \Eprint
  {http://arxiv.org/abs/http://dx.doi.org/10.1080/00268976.2015.1119902}
  {http://dx.doi.org/10.1080/00268976.2015.1119902} \BibitemShut {NoStop}%
\bibitem [{\citenamefont {Tasn\'{a}di}\ and\ \citenamefont
  {Nagy}(2003)}]{tasnadi_nagy2003}%
  \BibitemOpen
  \bibfield  {author} {\bibinfo {author} {\bibfnamefont {F.}~\bibnamefont
  {Tasn\'{a}di}}\ and\ \bibinfo {author} {\bibfnamefont {A.}~\bibnamefont
  {Nagy}},\ }\href@noop {} {\bibfield  {journal} {\bibinfo  {journal} {J. Phys.
  B: At. Mol. Opt. Phys.}\ }\textbf {\bibinfo {volume} {36}},\ \bibinfo {pages}
  {4073} (\bibinfo {year} {2003})}\BibitemShut {NoStop}%
\bibitem [{\citenamefont {Lieb}(1983)}]{LFTransform-Lieb}%
  \BibitemOpen
  \bibfield  {author} {\bibinfo {author} {\bibfnamefont {E.~H.}\ \bibnamefont
  {Lieb}},\ }\href@noop {} {\bibfield  {journal} {\bibinfo  {journal} {Int. J.
  Quantum Chem.}\ }\textbf {\bibinfo {volume} {24}},\ \bibinfo {pages} {243}
  (\bibinfo {year} {1983})}\BibitemShut {NoStop}%
\bibitem [{\citenamefont {Kvaal}\ \emph {et~al.}(2014)\citenamefont {Kvaal},
  \citenamefont {Ekstr\"{o}m}, \citenamefont {Teale},\ and\ \citenamefont
  {Helgaker}}]{differentiable2014}%
  \BibitemOpen
  \bibfield  {author} {\bibinfo {author} {\bibfnamefont {S.}~\bibnamefont
  {Kvaal}}, \bibinfo {author} {\bibfnamefont {U.}~\bibnamefont {Ekstr\"{o}m}},
  \bibinfo {author} {\bibfnamefont {A.~M.}\ \bibnamefont {Teale}}, \ and\
  \bibinfo {author} {\bibfnamefont {T.}~\bibnamefont {Helgaker}},\ }\href@noop
  {} {\bibfield  {journal} {\bibinfo  {journal} {J. Chem. Phys.}\ }\textbf
  {\bibinfo {volume} {140}},\ \bibinfo {pages} {18A518} (\bibinfo {year}
  {2014})}\BibitemShut {NoStop}%
\bibitem [{\citenamefont {Stoll}\ and\ \citenamefont
  {Savin}(1985)}]{savinstoll}%
  \BibitemOpen
  \bibfield  {author} {\bibinfo {author} {\bibfnamefont {H.}~\bibnamefont
  {Stoll}}\ and\ \bibinfo {author} {\bibfnamefont {A.}~\bibnamefont {Savin}},\
  }in\ \href@noop {} {\emph {\bibinfo {booktitle} {Density Functional Methods
  in Physics}}},\ \bibinfo {editor} {edited by\ \bibinfo {editor}
  {\bibfnamefont {R.~M.}\ \bibnamefont {Dreizler}}\ and\ \bibinfo {editor}
  {\bibfnamefont {J.}~\bibnamefont {da~Providencia}}}\ (\bibinfo  {publisher}
  {Plenum},\ \bibinfo {address} {New York},\ \bibinfo {year}
  {1985})\BibitemShut {NoStop}%
\bibitem [{\citenamefont {Savin}(1996)}]{savinbook}%
  \BibitemOpen
  \bibfield  {author} {\bibinfo {author} {\bibfnamefont {A.}~\bibnamefont
  {Savin}},\ }\href@noop {} {\emph {\bibinfo {title} {Recent Developments and
  Applications of Modern Density Functional Theory}}}\ (\bibinfo  {publisher}
  {Elsevier},\ \bibinfo {address} {Amsterdam},\ \bibinfo {year} {1996})\ p.\
  \bibinfo {pages} {327}\BibitemShut {NoStop}%
\bibitem [{\citenamefont {Savin}(1988)}]{savin1988combined}%
  \BibitemOpen
  \bibfield  {author} {\bibinfo {author} {\bibfnamefont {A.}~\bibnamefont
  {Savin}},\ }\href@noop {} {\bibfield  {journal} {\bibinfo  {journal} {Int. J.
  Quantum Chem.}\ }\textbf {\bibinfo {volume} {34}},\ \bibinfo {pages} {59}
  (\bibinfo {year} {1988})}\BibitemShut {NoStop}%
\bibitem [{\citenamefont {Levy}(1995)}]{PRA_Levy_XE-N-N-1}%
  \BibitemOpen
  \bibfield  {author} {\bibinfo {author} {\bibfnamefont {M.}~\bibnamefont
  {Levy}},\ }\href {\doibase 10.1103/PhysRevA.52.R4313} {\bibfield  {journal}
  {\bibinfo  {journal} {Phys. Rev. A}\ }\textbf {\bibinfo {volume} {52}},\
  \bibinfo {pages} {R4313} (\bibinfo {year} {1995})}\BibitemShut {NoStop}%
\bibitem [{\citenamefont {Toulouse}\ \emph {et~al.}(2005)\citenamefont
  {Toulouse}, \citenamefont {Gori-Giorgi},\ and\ \citenamefont
  {Savin}}]{Toulouse2005TCA}%
  \BibitemOpen
  \bibfield  {author} {\bibinfo {author} {\bibfnamefont {J.}~\bibnamefont
  {Toulouse}}, \bibinfo {author} {\bibfnamefont {P.}~\bibnamefont
  {Gori-Giorgi}}, \ and\ \bibinfo {author} {\bibfnamefont {A.}~\bibnamefont
  {Savin}},\ }\href@noop {} {\bibfield  {journal} {\bibinfo  {journal} {Theor.
  Chem. Acc.}\ }\textbf {\bibinfo {volume} {114}},\ \bibinfo {pages} {305}
  (\bibinfo {year} {2005})}\BibitemShut {NoStop}%
\bibitem [{\citenamefont {Gould}\ and\ \citenamefont
  {Dobson}(2013)}]{gould2013}%
  \BibitemOpen
  \bibfield  {author} {\bibinfo {author} {\bibfnamefont {T.}~\bibnamefont
  {Gould}}\ and\ \bibinfo {author} {\bibfnamefont {J.~F.}\ \bibnamefont
  {Dobson}},\ }\href@noop {} {\bibfield  {journal} {\bibinfo  {journal} {J.
  Chem. Phys.}\ }\textbf {\bibinfo {volume} {138}},\ \bibinfo {pages} {014103}
  (\bibinfo {year} {2013})}\BibitemShut {NoStop}%
\bibitem [{\citenamefont {Paziani}\ \emph {et~al.}(2006)\citenamefont
  {Paziani}, \citenamefont {Moroni}, \citenamefont {Gori-Giorgi},\ and\
  \citenamefont {Bachelet}}]{Paziani2006PRB}%
  \BibitemOpen
  \bibfield  {author} {\bibinfo {author} {\bibfnamefont {S.}~\bibnamefont
  {Paziani}}, \bibinfo {author} {\bibfnamefont {S.}~\bibnamefont {Moroni}},
  \bibinfo {author} {\bibfnamefont {P.}~\bibnamefont {Gori-Giorgi}}, \ and\
  \bibinfo {author} {\bibfnamefont {G.~B.}\ \bibnamefont {Bachelet}},\
  }\href@noop {} {\bibfield  {journal} {\bibinfo  {journal} {Phys. Rev. B}\
  }\textbf {\bibinfo {volume} {73}},\ \bibinfo {pages} {155111} (\bibinfo
  {year} {2006})}\BibitemShut {NoStop}%
\bibitem [{\citenamefont {Rebolini}\ \emph {et~al.}(2015)\citenamefont
  {Rebolini}, \citenamefont {Toulouse}, \citenamefont {Teale}, \citenamefont
  {Helgaker},\ and\ \citenamefont {Savin}}]{MP15_Elisa_PT_XE_AC}%
  \BibitemOpen
  \bibfield  {author} {\bibinfo {author} {\bibfnamefont {E.}~\bibnamefont
  {Rebolini}}, \bibinfo {author} {\bibfnamefont {J.}~\bibnamefont {Toulouse}},
  \bibinfo {author} {\bibfnamefont {A.~M.}\ \bibnamefont {Teale}}, \bibinfo
  {author} {\bibfnamefont {T.}~\bibnamefont {Helgaker}}, \ and\ \bibinfo
  {author} {\bibfnamefont {A.}~\bibnamefont {Savin}},\ }\href@noop {}
  {\bibfield  {journal} {\bibinfo  {journal} {Mol. Phys.}\ }\textbf {\bibinfo
  {volume} {113}},\ \bibinfo {pages} {1740} (\bibinfo {year}
  {2015})}\BibitemShut {NoStop}%
\bibitem [{\citenamefont {Stoyanova}\ \emph {et~al.}(2013)\citenamefont
  {Stoyanova}, \citenamefont {Teale}, \citenamefont {Toulouse}, \citenamefont
  {Helgaker},\ and\ \citenamefont {Fromager}}]{manusroep2013}%
  \BibitemOpen
  \bibfield  {author} {\bibinfo {author} {\bibfnamefont {A.}~\bibnamefont
  {Stoyanova}}, \bibinfo {author} {\bibfnamefont {A.~M.}\ \bibnamefont
  {Teale}}, \bibinfo {author} {\bibfnamefont {J.}~\bibnamefont {Toulouse}},
  \bibinfo {author} {\bibfnamefont {T.}~\bibnamefont {Helgaker}}, \ and\
  \bibinfo {author} {\bibfnamefont {E.}~\bibnamefont {Fromager}},\ }\href@noop
  {} {\bibfield  {journal} {\bibinfo  {journal} {J. Chem. Phys.}\ }\textbf
  {\bibinfo {volume} {139}},\ \bibinfo {pages} {134113} (\bibinfo {year}
  {2013})}\BibitemShut {NoStop}%
\bibitem [{\citenamefont {Toulouse}\ \emph {et~al.}(2004)\citenamefont
  {Toulouse}, \citenamefont {Savin},\ and\ \citenamefont {Flad}}]{toulda}%
  \BibitemOpen
  \bibfield  {author} {\bibinfo {author} {\bibfnamefont {J.}~\bibnamefont
  {Toulouse}}, \bibinfo {author} {\bibfnamefont {A.}~\bibnamefont {Savin}}, \
  and\ \bibinfo {author} {\bibfnamefont {H.~J.}\ \bibnamefont {Flad}},\
  }\href@noop {} {\bibfield  {journal} {\bibinfo  {journal} {Int. J. Quantum
  Chem.}\ }\textbf {\bibinfo {volume} {100}},\ \bibinfo {pages} {1047}
  (\bibinfo {year} {2004})}\BibitemShut {NoStop}%
\bibitem [{\citenamefont {Savin}(2014)}]{savin2014towards}%
  \BibitemOpen
  \bibfield  {author} {\bibinfo {author} {\bibfnamefont {A.}~\bibnamefont
  {Savin}},\ }\href@noop {} {\bibfield  {journal} {\bibinfo  {journal} {J.
  Chem. Phys.}\ }\textbf {\bibinfo {volume} {140}},\ \bibinfo {pages} {18A509}
  (\bibinfo {year} {2014})}\BibitemShut {NoStop}%
\bibitem [{\citenamefont {Cornaton}\ \emph {et~al.}(2013)\citenamefont
  {Cornaton}, \citenamefont {Stoyanova}, \citenamefont {Jensen},\ and\
  \citenamefont {Fromager}}]{Cornaton2013PRA}%
  \BibitemOpen
  \bibfield  {author} {\bibinfo {author} {\bibfnamefont {Y.}~\bibnamefont
  {Cornaton}}, \bibinfo {author} {\bibfnamefont {A.}~\bibnamefont {Stoyanova}},
  \bibinfo {author} {\bibfnamefont {H.~J.~{\Aa}.}\ \bibnamefont {Jensen}}, \
  and\ \bibinfo {author} {\bibfnamefont {E.}~\bibnamefont {Fromager}},\
  }\href@noop {} {\bibfield  {journal} {\bibinfo  {journal} {Phys. Rev. A}\
  }\textbf {\bibinfo {volume} {88}},\ \bibinfo {pages} {022516} (\bibinfo
  {year} {2013})}\BibitemShut {NoStop}%
\bibitem [{DAL()}]{DALTON}%
  \BibitemOpen
  \href@noop {} {\enquote {\bibinfo {title} {Dalton, a molecular electronic
  structure program, release dalton2015 (2015), see
  http://daltonprogram.org/},}\ }\BibitemShut {NoStop}%
\bibitem [{\citenamefont {Aidas}\ \emph {et~al.}(2015)\citenamefont {Aidas}
  \emph {et~al.}}]{DALTON_short}%
  \BibitemOpen
  \bibfield  {author} {\bibinfo {author} {\bibfnamefont {K.}~\bibnamefont
  {Aidas}} \emph {et~al.},\ }\href {\doibase 10.1002/wcms.1172} {\bibfield
  {journal} {\bibinfo  {journal} {WIREs Comput.~Mol.~Sci.}\ }\textbf {\bibinfo
  {volume} {4}},\ \bibinfo {pages} {269} (\bibinfo {year} {2015})}\BibitemShut
  {NoStop}%
\bibitem [{\citenamefont {Dunning~Jr}(1989)}]{dunning1989gaussian}%
  \BibitemOpen
  \bibfield  {author} {\bibinfo {author} {\bibfnamefont {T.~H.}\ \bibnamefont
  {Dunning~Jr}},\ }\href@noop {} {\bibfield  {journal} {\bibinfo  {journal} {J.
  Chem. Phys.}\ }\textbf {\bibinfo {volume} {90}},\ \bibinfo {pages} {1007}
  (\bibinfo {year} {1989})}\BibitemShut {NoStop}%
\bibitem [{\citenamefont {Woon}\ and\ \citenamefont
  {Dunning~Jr}(1994)}]{woon1994gaussian}%
  \BibitemOpen
  \bibfield  {author} {\bibinfo {author} {\bibfnamefont {D.~E.}\ \bibnamefont
  {Woon}}\ and\ \bibinfo {author} {\bibfnamefont {T.~H.}\ \bibnamefont
  {Dunning~Jr}},\ }\href@noop {} {\bibfield  {journal} {\bibinfo  {journal} {J.
  Chem. Phys.}\ }\textbf {\bibinfo {volume} {100}},\ \bibinfo {pages} {2975}
  (\bibinfo {year} {1994})}\BibitemShut {NoStop}%
\bibitem [{\citenamefont {Franck}\ \emph {et~al.}(2015)\citenamefont {Franck},
  \citenamefont {Mussard}, \citenamefont {Luppi},\ and\ \citenamefont
  {Toulouse}}]{JCP15_Odile_basis_convergence_srDFT}%
  \BibitemOpen
  \bibfield  {author} {\bibinfo {author} {\bibfnamefont {O.}~\bibnamefont
  {Franck}}, \bibinfo {author} {\bibfnamefont {B.}~\bibnamefont {Mussard}},
  \bibinfo {author} {\bibfnamefont {E.}~\bibnamefont {Luppi}}, \ and\ \bibinfo
  {author} {\bibfnamefont {J.}~\bibnamefont {Toulouse}},\ }\href {\doibase
  http://dx.doi.org/10.1063/1.4907920} {\bibfield  {journal} {\bibinfo
  {journal} {J. Chem. Phys.}\ }\textbf {\bibinfo {volume} {142}},\ \bibinfo
  {eid} {074107} (\bibinfo {year} {2015})}\BibitemShut {NoStop}%
\end{thebibliography}

\newcommand{\Aa}[0]{Aa}
%


\end{document}